\documentclass[aps,prx,twocolumn,superscriptaddress,showpacs,amsmath,amssymb]
{revtex4-1}
\usepackage{booktabs}%\documentclass[aps,prl,preprint,superscriptaddress]{revtex4-2}
\usepackage{graphicx}
\usepackage{subfigure}
\usepackage{gensymb}
\usepackage{tabularx}
\usepackage{ltablex}
\usepackage{nicefrac}
\begin{document}

	%\preprint{}
	\title{Surface structure and stacking of the commensurate\\ ($\sqrt{13} \times
		\sqrt{13}$)R13.9$^\circ$  charge density wave phase of
		1\textit{T}-TaS$_2$(0001)}
	
	
	\author{Gevin von Witte}
	\affiliation{IV. Physical Institute, Georg-August-University
		G\"{o}ttingen, D-37077 G\"{o}ttingen, Germany}
	
	\author{Tilman Ki{\ss}linger} \affiliation{Solid
		State Physics, Friedrich-Alexander-University Erlangen-N\"{u}rnberg,
		D-91058 Erlangen, Germany}
	
	\author{Jan Gerrit Horstmann}
	\affiliation{IV. Physical Institute, Georg-August-University
		G\"{o}ttingen, D-37077 G\"{o}ttingen, Germany}
	
	\author{\mbox{Kai Rossnagel}}
	\affiliation{Institut f\"ur Experimentelle und Angewandte Physik, Christian-Albrechts-Universit\"at zu Kiel, D-24098 Kiel, Germany}
	\affiliation{Ruprecht-Haensel-Labor, Christian-Albrechts-Universit\"at zu Kiel und Deutsches Elektronen-Synchrotron DESY, D-24098 Kiel und D-22607 Hamburg, Germany}
	\affiliation{Deutsches Elektronen-Synchrotron DESY, D-22607 Hamburg}
	
	\author{M. Alexander Schneider} \affiliation{Solid
		State Physics, Friedrich-Alexander-University Erlangen-N\"{u}rnberg,
		D-91058 Erlangen, Germany}
	
	\author{Claus Ropers}
	\affiliation{IV. Physical Institute, Georg-August-University
		G\"{o}ttingen, D-37077 G\"{o}ttingen, Germany}
	
	\author{Lutz Hammer}
	\email{lutz.hammer@fau.de} \affiliation{Solid State Physics,
		Friedrich-Alexander-University Erlangen-N\"{u}rnberg, D-91058
		Erlangen, Germany}
	%\thanks{}
	

\begin{abstract}
\vspace{0.5cm}

By quantitative low-energy electron diffraction (LEED) we investigate the extensively studied 
commensurate charge density wave (CDW) phase of trigonal tantalum disulphide (1\textit{T}-TaS$_2$), 
which develops at low temperatures with a ($\sqrt{13} \times\sqrt{13}$)R13.9$^\circ$ periodicity. 
A full-dynamical analysis of the energy dependence of diffraction spot intensities reveals 
the entire crystallographic surface structure, i.e. the detailed atomic
positions within the outermost two trilayers consisting of 78 atoms as well as the CDW
stacking. The analysis is based on an unusually large data set consisting of spectra for
128 inequivalent beams taken in the energy range 20 - 250\,eV and an
excellent fit quality expressed by a bestfit Pendry R-factor of R = 0.110.

The LEED intensity analysis reveals that the well-accepted model of
star-of-David-shaped clusters of Ta atoms for the bulk
structure also holds for the outermost two TaS$_2$ trilayers.
Specifically, in both layers the clusters of Ta atoms contract
laterally by up to 0.25\,\AA\ and also slightly rotate within the
superstructure cell, causing respective distortions as well as heavy
bucklings (up to 0.23\,\AA) in the adjacent sulphur layers. Most
importantly, our analysis finds that the CDWs of the 1$^{st}$ and
2$^{nd}$ trilayer are vertically aligned, while there is a lateral
shift of two units of the basic hexagonal lattice (6.71\,\AA)
between the 2$^{nd}$ and 3$^{rd}$ trilayer.

The results may contribute to a better understanding of the intricate electronic structure 
of the reference compound 1\textit{T}-TaS2 and guide the way to the analysis of complex structures in similar quantum materials.

\end{abstract}


% 61.05.jh    Low-energy electron diffraction (LEED) and reflection high-energy electron diffraction (RHEED)
% 68.35.B-    Structure of clean surfaces and surface reconstruction
% 71.45.Lr    Charge-density-wave systems

\maketitle

%\subsection{}
%\subsubsection{}

\section{Introduction}
The transition-metal dichalcogenide 1$T$-TaS$_2$ consists of
van der Waals-stacked S--Ta--S trilayers. The 1$T$ polytype with
octahedrally coordinated Ta atoms exhibits one commensurate
$(\sqrt{13} \times \sqrt{13}$)R13.9$^\circ$ charge-density wave
phase (C-phase, $T < 187$\,K) and other non-commensurate CDW states
at higher temperatures (187\,K - 543\,K) \cite{Scruby1975}. Having
been known for decades, these different phases and their transitions 
\cite{Vogelgesang2018} are experiencing renewed and growing
interest, e.g., following the discovery of pressure-induced 
superconductivity \cite{Sipos2008} and optical and 
electrical switching to metastable ``hidden'' CDW states  
\cite{Stojchevska2014,Cho2016,Vaskivskyi2016,Ma2016,Gerasimenko2019}, 
the observation of ultrafast electronic structure changes at the surface 
\cite{Perfetti2006,Petersen2011,Hellmann2012,Ligges2018} 
and trilayer number-dependent CDW phases in thin crystals \cite{Yu2015},
as well as the prediction of complex orbital textures \cite{Ritschel2015} 
and a quantum spin-liquid state associated with the C-phase \cite{Law2017,Klanjsek2017}.
Whereas the electronic properties of a single trilayer may often be a good starting point 
to explain these various phenomena \cite{Rossnagel2006}, it has recently become clear 
that a full understanding of the electronic structure of the CDW states requires 
to include interlayer coupling and the stacking order of the CDW perpendicular 
to the trilayers \cite{Ritschel2015}.
In the C-phase of 1$T$-TaS$_2$, for example, 
different CDW stackings can result in metallic or insulating behavior 
within the trilayers, but also in a state that corresponds 
to a band insulator within the trilayers and a metal 
in the perpendicular direction \cite{Darancet2014,Ritschel2018,Ngankeu2017,Lee2019}.
Thus, detailed knowledge of the three-dimensional atomic and electronic CDW structure 
is essential, and particularly its modification at the surface 
as many of the recent key experimental observations were made by surface-sensitive techniques 
such as scanning tunneling microscopy \cite{Cho2016,Ma2016,Gerasimenko2019}, 
time- and angle-resolved photoemission spectroscopy 
\cite{Perfetti2006,Petersen2011,Hellmann2012,Ligges2018}, or ultrafast LEED \cite{Vogelgesang2018}.

Concerning the atomic structure of the CDW phases, Brouwer and 
Jellinek \cite{Brouwer1980} have taken room-temperature X-ray
diffraction (XRD) data for the so-called nearly commensurate
(NC) phase, but evaluated them in the approximation of the
low-temperature commensurate structure. The true domain structure of
the NC-phase has been analyzed in a later XRD study by
Yamamoto \cite{Yamamoto1983} and refined by Spijkerman 
et al. \cite{Spijkerman1997}. Thus, strictly speaking, for the
low-temperature C-phase not even the bulk structure has been precisely
determined so far. Only the CDW stacking has been revealed by cross-section transmission electron microscopy, where an alternation of vertical stacking and a particular type of side-shift vectors was found \cite{Ishiguro1991}. In this contribution, we investigate the surface structure of this C-phase by a quantitative LEED intensity analysis.
The small penetration depth of low-energy electrons yields 
quantitative structural information for the first and also 
- with somewhat less accuracy - the second trilayer, 
which may be taken as representative of the bulk structure. 

Recently, Chen and coworkers \cite{Chen2018} used quantitative LEED 
to determine the surface structure of the 1$T$-TaTe$_2$(0001)-(3$\times$1) phase, 
a closely related but less complicated transition metal dichalcogenide phase. Based on a
bestfit Pendry R-factor \cite{Pendry} value of R = 0.29, they revealed an 
almost bulk-like structure for the outermost TaTe$_2$ trilayer. In the present investigation, 
we will show that a much lower reliability factor value (here $R = 0.11$) is necessary 
to unambiguously reveal the surface structure of the more complex ($\sqrt{13} \times \sqrt{13}$)R13.9$^\circ$ phase of 1\textit{T}-TaS$_2(0001)$ and to characterize subtle deviations from the bulk-like structure. Such a fit quality is by no means standard for LEED analyses of systems with such large unit cells. Therefore, this study also intends to provide guidance for the analysis of complex structures through precise data acquisition and consideration of all relevant structural and non-structural parameters in the intensity calculations.

\section{LEED experiment and intensity calculations}

\subsection{LEED experiments} \label{Experiment}

All experiments were performed in a standard ultra-high vacuum (UHV) chamber equipped with all necessary facilities for sample handling, LEED and scanning tunneling microscopy (STM). Details of the apparatus, the sample cleavage and the STM characterization of the surface morphology are given in Appendix\,\ref{ExpDetails}.  

\begin{figure}[htb]
	\begin{centering}
		\includegraphics[width=0.45\textwidth]{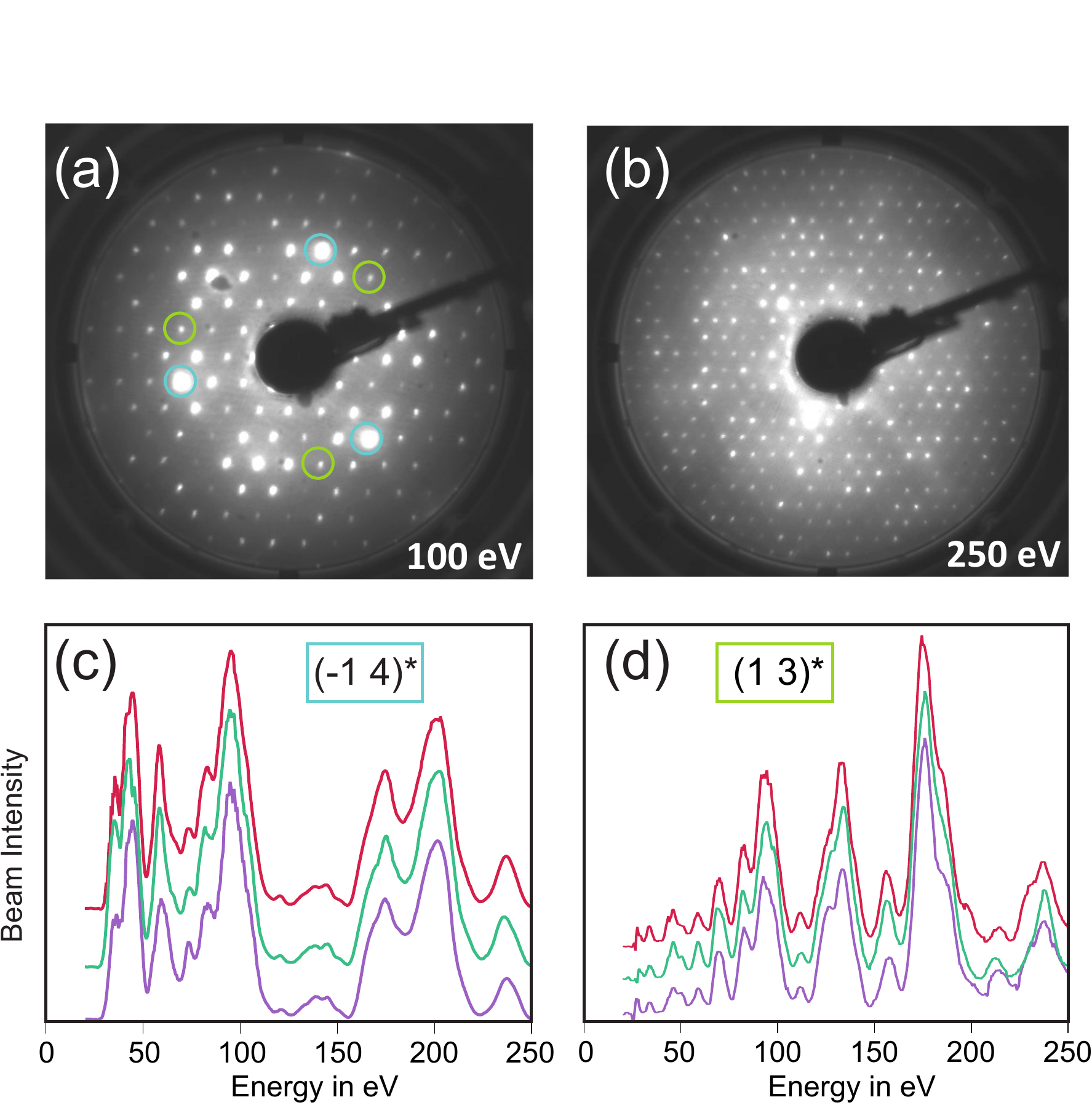}
		\caption{(Color online) (a,b): LEED pattern of the ($\sqrt{13} \times
			\sqrt{13}$)R13.9$^\circ$ phase for different energies. (c,d): 
			Comparison of raw experimental IV-spectra for beam triples indicated
			in the LEED image (a) linked by a 120$^\circ$ rotation. For beam
			labeling see text.}\label{LEEDSymmetry}
	\end{centering}
\end{figure}

The freshly \textit{in-situ} cleaved TaS$_2$ sample held at about 100~K showed the
diffraction pattern of a well-ordered and mono-domain ($\sqrt{13}
\times \sqrt{13}$)R13.9$^\circ$ CDW phase, which is displayed for 
two energies in Fig.~\ref{LEEDSymmetry}(a,b). 
No sign of the second symmetry-equivalent domain rotated 
by -13.9$^\circ$ against the basic lattice 
(cf. Fig.\,\ref{Nomenclature}(a) could be detected across the whole crystal.
Right after cleavage, the TaS$_2$ surface was aligned in front of
the LEED optics for normal incidence of the electron beam 
via maximum concordance of the intensity vs. energy (voltage) characteristics (IV-spectra)
for selected beam triples linked by a 120$^\circ$ rotation (cf.
Fig.\,\ref{LEEDSymmetry}(c,d)). Directly thereafter, a series of images of
the complete diffraction pattern was taken as a function of energy
(20-250\,eV in steps of 0.5\,eV) by a highly sensitive CCD camera
and stored on a computer. The whole LEED data acquisition including initial
alignment was done within less than one hour after cleavage, i.e.,
surface contamination can be assumed to be negligible.

Several cycles of heating to $\sim$400\,K 
into the IC phase (higher temperatures were incompatible with
the adhesive crystal mounting) and subsequent cooling did not lead
to any visible changes in the LEED pattern or within the IV
spectra. With the conversion into the IC phase, both the local
trilayer structure and the CDW stacking are changed
completely \cite{Scruby1975}. Hence the full structural restoration
upon cooling, verified by identical LEED-IV spectra, can be taken as a
proof that our analysis indeed describes the \emph{equilibrium}
surface structure of the C-phase of 1\textit{T}-TaS$_2(0001)$.

A visual inspection of the LEED patterns in Fig.~\ref{LEEDSymmetry} 
already suggests a 3-fold symmetry, which becomes even more clearly 
by comparing IV-spectra for beams being related by a 120$^\circ$ rotation. 
As shown for two examples in the lower part of the figure all three
spectra look rather identical. Of course there are tiny variations
in the raw data for the different beams, however, the size of these
differences are in the typical range of experimental misalignment in
particular of the normal incidence.

The stored stack of LEED images made up the raw data base for LEED
intensity spectra. In an offline evaluation the IV-spectra for all accessible
beams (370 in total) were extracted  using the programme package 
\emph{EE2010} \cite{EE2010}, which automatically corrects 
for the background of the quasi-elastically scattered electrons. 
This automatic correction requires sufficient
space between neighboring diffraction spots to avoid any crosstalk. 
Thus, the evaluable energy range is limited by the spot density
(cf. Fig.~\ref{LEEDSymmetry}(b)).

The post-processing of the spectra involved the normalization to the
measured, energy-dependent primary beam intensity, averaging over
symmetry-related beams, correction for the cosine of the viewing
angle as well as noise filtering (7-point median and four times
4$^{th}$ order Savitzky-Golay smoothing over 27 points). This
procedure finally resulted in a total data base of 128 symmetrically
independent experimental beams for the LEED analysis with a
cumulated energy width of $\Delta E = 15383$~eV. The complete set of
experimental data used as input for the fitting procedure is
provided in the Supplemental Material \cite{SupMat}.

\subsection{LEED-IV calculations and error determination} \label{LEEDcalc}
Full-dynamical LEED intensity calculations for selected model configurations 
were performed using the Erlangen LEED code \emph{TensErLEED} \cite{tenserleed}. 
Starting from these reference structures we conducted an extended fitting procedure 
using the perturbation method Tensor LEED \cite{Heinz1995,Rous1986} in combination with a
modified random sampling search algorithm \cite{Kottcke1997}, both of which 
are also implemented in the \emph{TensErLEED} program package. 
After each fitting cycle the resulting  bestfit structure was verified 
by a new full-dynamical intensity calculation, 
which then served as the reference structure for the next run. 
Details of the computation are given in Appendix\,\ref{CompDetails}.

\begin{figure}[htb]
	\begin{centering}
		\includegraphics[width=0.45\textwidth]{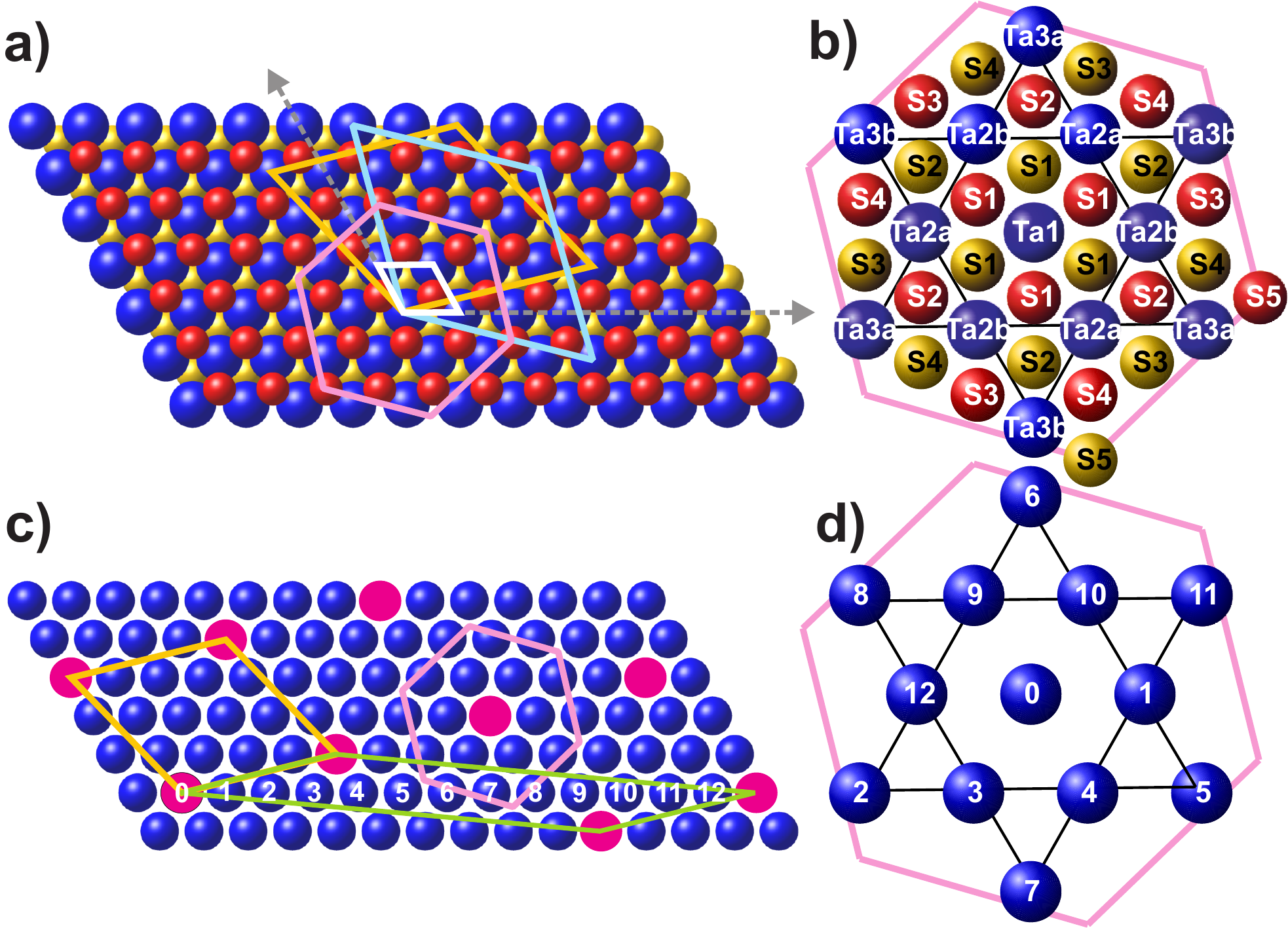}
		\caption{(Color online) (a) Ball model of a TaS$_2$ trilayer with
			the two symmetry equivalent unit meshes of the ($\sqrt{13} \times
			\sqrt{13}$)R13.9$^\circ$ CDW phase (yellow, cyan) and that of the
			unreconstructed (1$\times$1) trilayer (white). Also shown in light purple is
			the hexagonal Wigner-Seitz (WS) cell for the yellow unit
			cell. (b) Labelling of Ta and S sites within the trilayer WS cell.
			(c) Alternative CDW unit cell (light green) illustrating the 13 different
			registry positions, where the central Ta atoms of the next (higher)
			trilayer can sit. (d) Projection of the registry positions into the
			WS cell.}\label{Nomenclature}
	\end{centering}
\end{figure}

In the course of the fit all geometrical parameters of the first (surface)
trilayer as well as those of lower lying trilayers (treated as
identical) were varied. As outlined in detail in Appendix\,\ref{Stacking}, 
this led to 5 inequivalent Ta and 10 S atoms per trilayer, which are labeled 
in Fig.\,\ref{Nomenclature}(b). So, we had to determine the geometrical positions 
of 30 atoms in total. Adapted to the assumed 3-fold symmetry of the trilayers, the
positional deviations were split up in parallel, radial or
azimuthal (rotational) displacements with respect to the vertical
rotation axis, which led to a total of 78 independent geometrical
parameters. 

Apart from the local atomic positions within each trilayer 
also the vertical stacking of the `$\sqrt{13}$\,'-supercells 
in surface-near trilayers had to be determined as well. In general,
there are as many different possible registries as atoms in the unit cell, 
see Fig.\,\ref{Nomenclature}(c). Because of the threefold rotational symmetry 
of the trilayers, the various stackings can be grouped again into 5 classes 
of symmetrically inequivalent configurations, 
which are \{0\}, \{1,3,9\}, \{4,10,12\}, \{2,5,6\}, and \{7,8,11\}
according to the labels given in Fig.~\ref{Nomenclature}(d). 

Despite the large number of determined structural and non-structural
parameters (86 in total) by this analysis, the huge collected
database ensures a large redundancy factor of $\varrho = \Delta E
/(N \cdot 4 V_\text{0i}) \approx 10$, i.e., we still have ten times
more data points than determined parameter values N. The degree of
correspondence between calculated and experimental spectra was
quantified by the Pendry R-factor $R$ \cite{Pendry}, which also
allows an estimate of the statistical errors of each parameter via
its variance $var(R)= R_\text{min}\cdot\sqrt{8\cdot
\text{V}_\text{0i}/\Delta E}$, see Section~\ref{Reliability}.

\section{Crystallographic structure of the 1\textit{T}-T\MakeLowercase{a}S$_2$($\sqrt{13} \times
\sqrt{13}$)R13.9$^\circ$ phase}

\subsection{Bestfit structural model} \label{Bestfit}

The starting point of the whole fitting procedure was the 
star-of-David-shaped reconstruction model with the parameter set
found by Brouwer and Jellinek for their room-temperature approximation of the structure \cite{Brouwer1980}. 
We interpreted the lateral shifts, mentioned but not further specified 
in that work, as being entirely radial with respect 
to the central Ta atom and assumed vertical stacking of the `$\sqrt{13}$\,'-supercells. 
Without any parameter refinement this structural model produced an only moderate
accordance of experimental and calculated data sets expressed by a
Pendry R-factor of $R = 0.326$. Such values are only sufficient to exclude
any stacking fault in the basic lattice at least within or right
below the first trilayer, since those models always led to R-factor
values above $R = 0.5$. We refrained from testing fundamentally different structural models but concentrated on the variation of local atomic positions as well as
on the superstructure stacking, since the moderate R-factor nevertheless gives some confidence that the star-of-David model also holds for the surface regime, although with local relaxations.

A change of the superstructure stacking implies that the
reconstruction pattern of (at least) one trilayer is shifted by one
or several basis vectors of the unreconstructed TaS$_2$ lattice.
With such a shift also the assignment of every single atom to
certain symmetry-related sites changes (cf.
Fig.~\ref{Nomenclature}(b)). Hence every stacking sequence has to be
treated as a separate structural model for which the whole set of
positional parameters has to be adjusted in an independent, complete fit. 
Since every non-vertical stacking breaks the 3-fold rotational symmetry 
of the overall system (for a detailed discussion see Appendix~\ref{Stacking}), 
a domain mixture has to be performed for those models. 
This means that the IV-spectra of beam triples related by threefold rotation 
have to be calculated separately and averaged prior to comparison 
with the experimental data.

In a series of fitting cycles we sequentially investigated 
all possible models for the CDW stacking of the first and 
also the second trilayer, which is described in detail 
in Appendix \ref{Fitting}. Eventually, we ended up 
with a bestfit model, where the stacking between 1$^{st}$ and 2$^{nd}$ trilayer 
is purely vertical (stacking class \{0\}), 
while that between 2$^{nd}$ and 3$^{rd}$ trilayer includes a side-shift 
by two unit vectors of the basic TaS$_2$ lattice (stacking class \{7,8,11\}). 
By that we find exactly the same alternating stacking sequence at the surface as
known for the bulk \cite{Ishiguro1991}, cf. Appendix \ref{Stacking}. 
All other stacking models could be excluded on the basis 
of the Pendry R-factor's variance. The bestfit model is characterized 
by an ultimate R-factor value of $R = 0.110$, which to our knowledge is
the lowest R-factor ever achieved for a LEED-IV analysis of this complexity.

\begin{figure}[htb]
	\begin{centering}
		\includegraphics[width=0.48\textwidth]{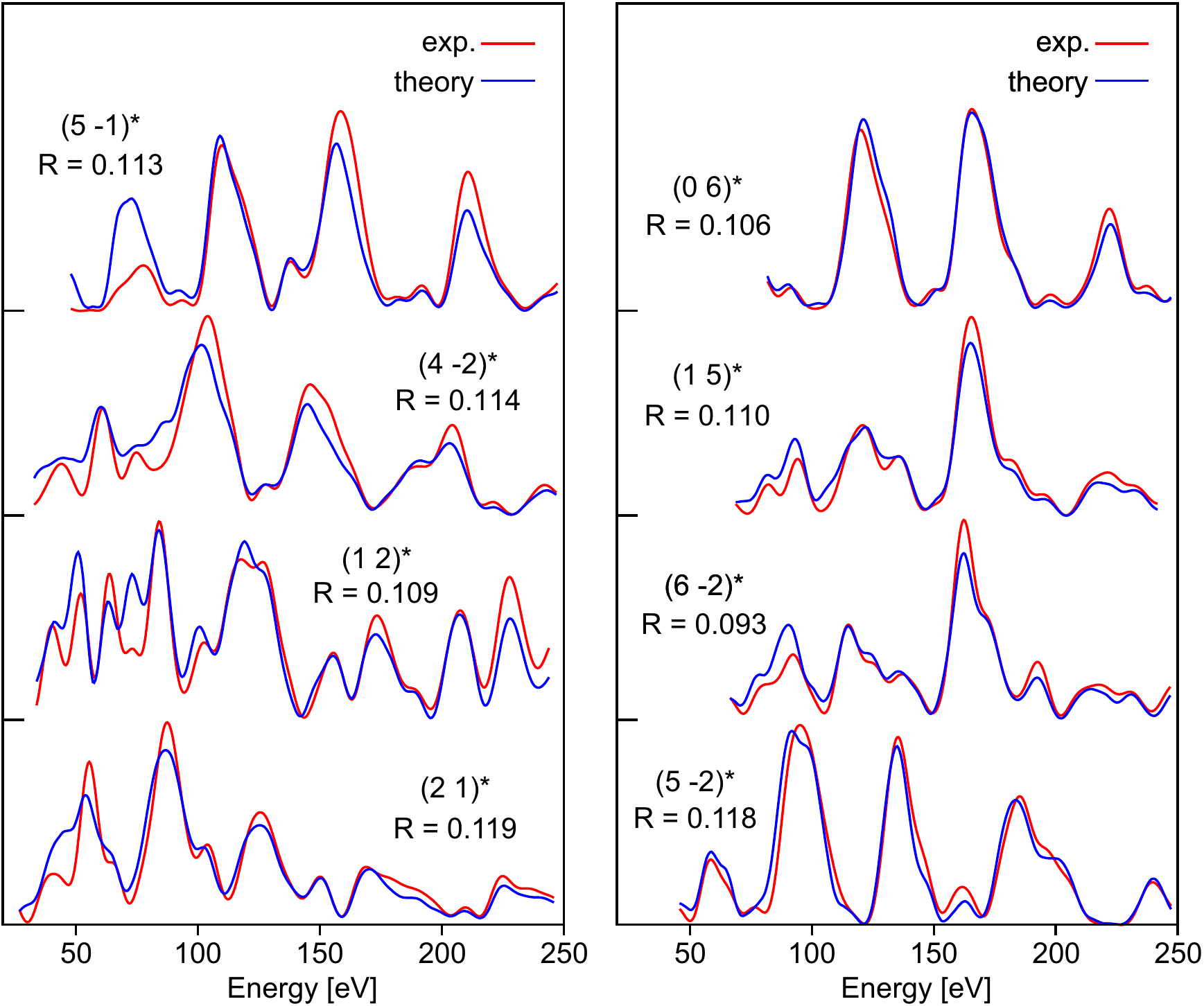}
		\caption{(Color online) Selection of experimental and bestfit
			IV-Spectra with single beam R-factors close to the overall R-factor
			R\,=\,0.110 of the analysis. For beam labeling see
			text.}\label{Spectra}
	\end{centering}
\end{figure}

The fit quality is visualized in Fig.\,\ref{Spectra} by a selection
of experimental and bestfit spectra with single beam R-factors close
to the overall R-factor of the analysis. A compilation of the full
set of 128 fitted beams is provided in the Supplemental Material 
\cite{SupMat}. For better readability, the beams are denoted as
multiples of the reciprocal basis vectors of the \emph{supercell
mesh}. For differention all reciprocal coordinates given according to
this superstructure basis are marked by an asterisk. The translation
into the usual coordinate system referring to the reciprocal lattice
of the unreconstructed TaS$_2$-crystal is given by $(3 ~~ 1)^*
\equiv (1 ~~ 0)$ and $(-1 ~~ 4)^* \equiv (0 ~~ 1)$ .

\begin{figure}[tbh]
\begin{centering}
\includegraphics[width=0.35\textwidth]{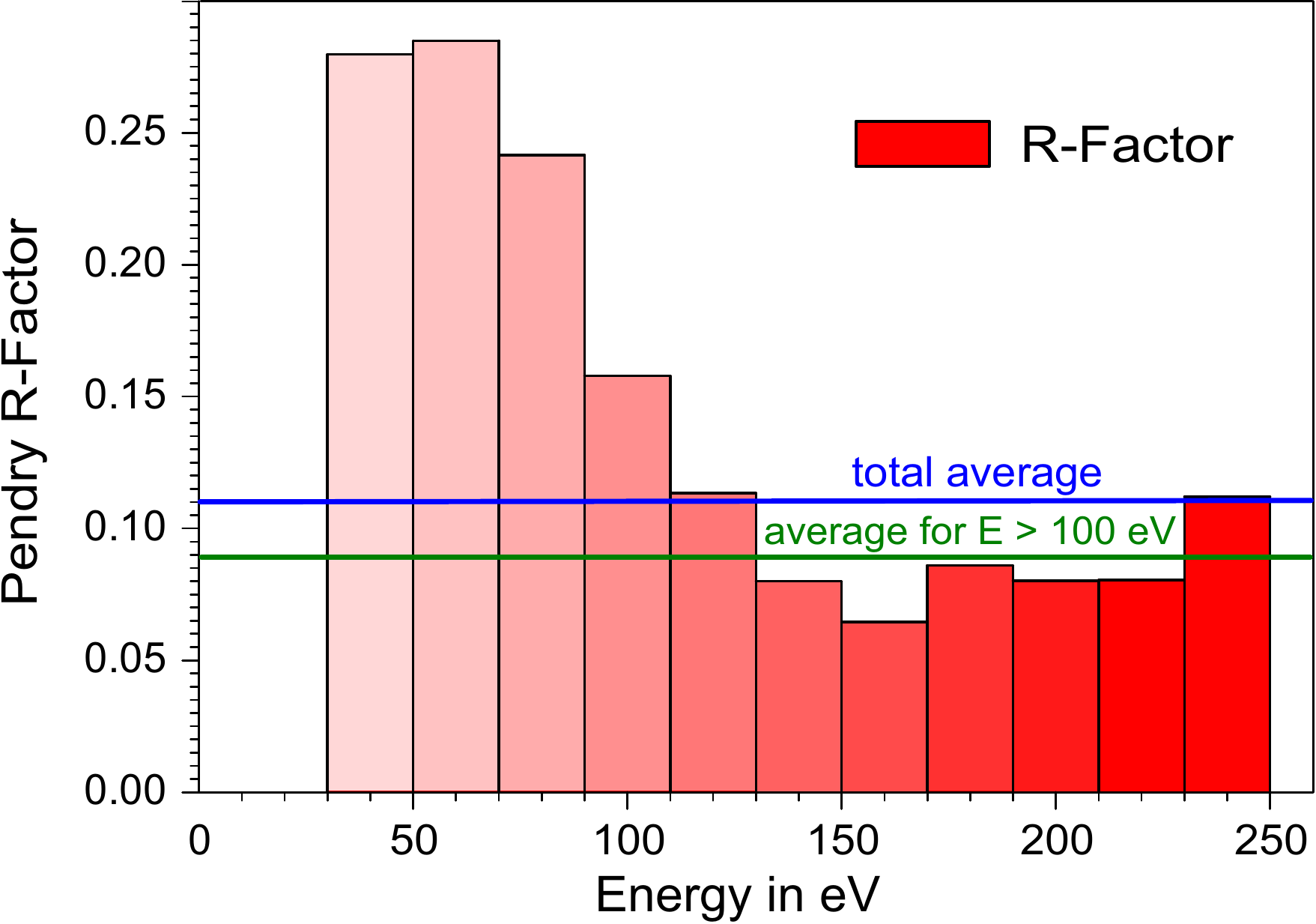}
\caption{(Color online) R-factor values calculated within the 
displayed energy increments of 20\,eV width.
Also shown is the overall R-factor level and that for energies above
100\,eV. The colour density of the columns correspond to the
statistical weight of the respective energy range.}\label{RvsE}
\end{centering}
\end{figure}

By careful examination of the spectra, it becomes obvious that the
fit quality is somewhat deteriorated in the very low energy range
below 100\,eV. This impression is reinforced by calculating the
overall R-factor within small energy intervals. This analysis 
is displayed in Fig.\,\ref{RvsE}, whereby the statistical weight 
of the respective intervals (i.e., their share of the total data base) is given by the colour density of the columns.
While above 100\,eV the bestfit R-factor is even as low as $R = 0.089$ 
on average, it rises up to about $R = 0.3$ for the lowest energies. 
Such an energy dependence of the R-factor is frequently
observed in particular when heavy atoms (like Ta in our case) are involved 
and is discussed for the case of Pt by
Materer et al. \cite{Materer1995}. They interpret this finding
as due to the incomplete consideration of spin effects in the LEED
computation by just using spin-averaged phase shifts and suggest to
ignore intensities below 100\,eV in the analysis. Here, we decided
to keep the full data basis for completeness.

\begin{figure}[htb]
	\begin{centering}
		\includegraphics[width=0.45\textwidth]{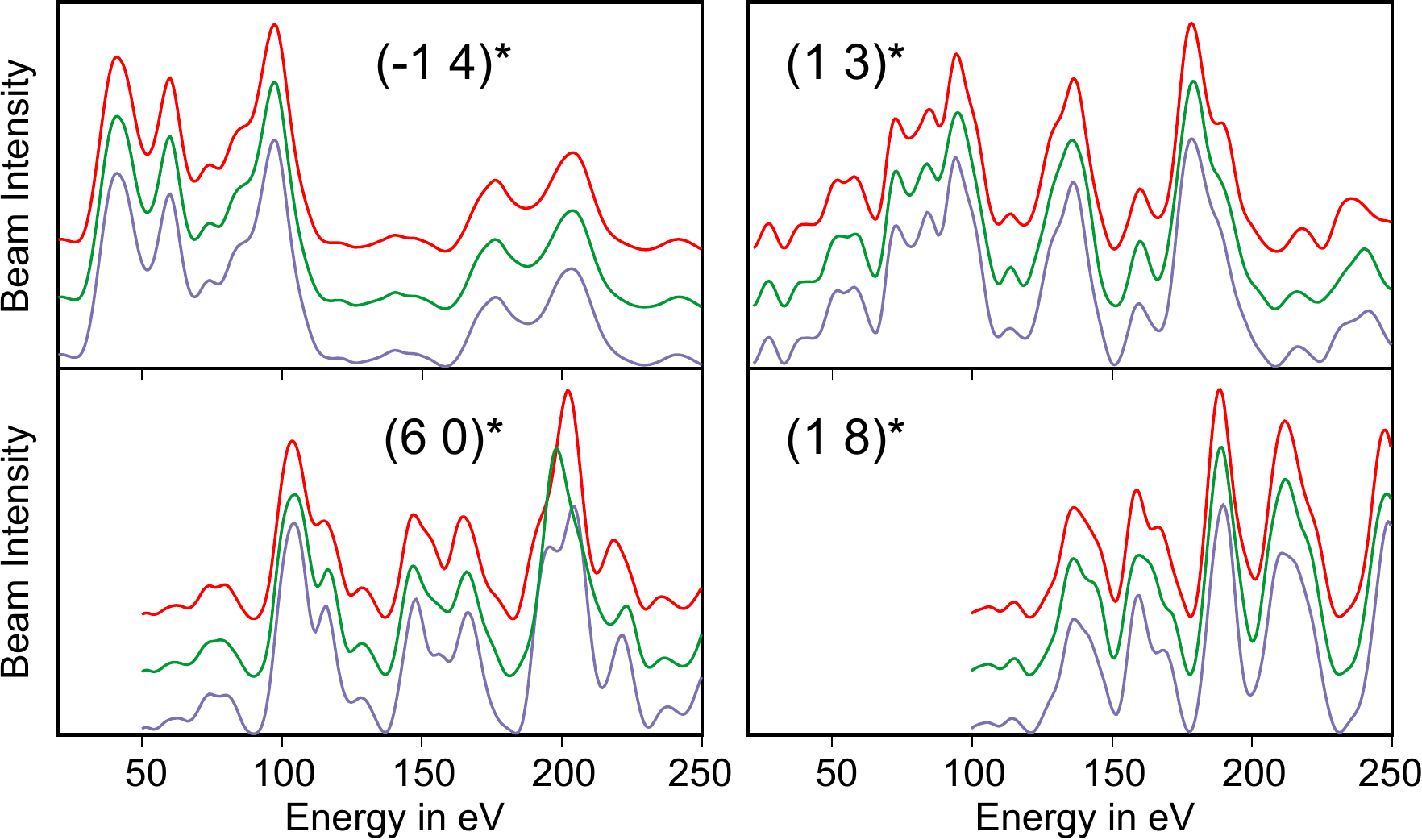}
		
		\caption{(Color online) Comparison of calculated spectra for beam
			triples linked by a 120$^\circ$ rotation.}\label{TheoSpectra}
	\end{centering}
\end{figure}

The alternate CDW stacking sequence \{0\} - \{7,8,11\} revealed by our LEED-IV analysis obviously breaks the 3-fold symmetry of the single trilayers 
and leaves the surface without any symmetry element (cf. Appendix~\ref{Stacking}).
At first glance this seems to be at variance with the experimental finding of a
3-fold symmetric LEED pattern (Fig.~\ref{LEEDSymmetry}). 
With the bestfit IV-spectra at hand, however, we can easily see 
that for the present structure this symmetry break is indeed very small. 
Fig.\,\ref{TheoSpectra} displays various triples of symmetry-related IV-spectra 
calculated for the bestfit structure. The upper row shows the counterparts 
to the experimental beams presented in Fig.\,\ref{LEEDSymmetry}(c.d), 
while the lower row displays two triples with maximum mutual spectral differences. 
Even in these examples the variations were smaller between calculated
spectra than towards the experimental ones. The smallness of the effect 
comes by two different reasons: First, the symmetry break 
occurs below the 6$^{th}$ layer, i.e., 12\,\AA\ deep in the crystal, 
where the electron wave-field is strongly attenuated. 
Second, the ascertained registry shift leads to a lateral position in space
being rather close to the next rotation axis within the WZ cell, 
which is at its corner, see Fig.~\ref{Nomenclature}(d). 
Another consequence is that we cannot discriminate from our LEED measurements, 
whether the determined lateral shift was indeed mono-domain, i.e., 
either \{7\}, \{8\}, or \{11\}. Alternatively, a domain mixture 
caused by possible domain boundaries within the CDW or surface steps 
cannot be ruled out. Accordingly, we cannot ascertain from our experiments, 
whether the system returns into exactly the same registry after cool-down 
from the IC-phase or switches between the three alternatives 
of the \{7,8,11\} stacking class.

\subsection{Reliability of the analysis} \label{Reliability}

Before we discuss the atomic structure and possible surface-induced
structural variations revealed by the LEED analysis, we first need to
know the precision with which the parameter values are determined. A
quantitative estimate of error margins is provided by the
reliability level of the R-factor $R + var(R)$ \cite{Pendry}:
All values for a certain parameter, which lead to a R-factor
below this level are assigned to belong to the error margin. 

\begin{figure}[htb]
\begin{centering}
\includegraphics[width=0.45\textwidth]{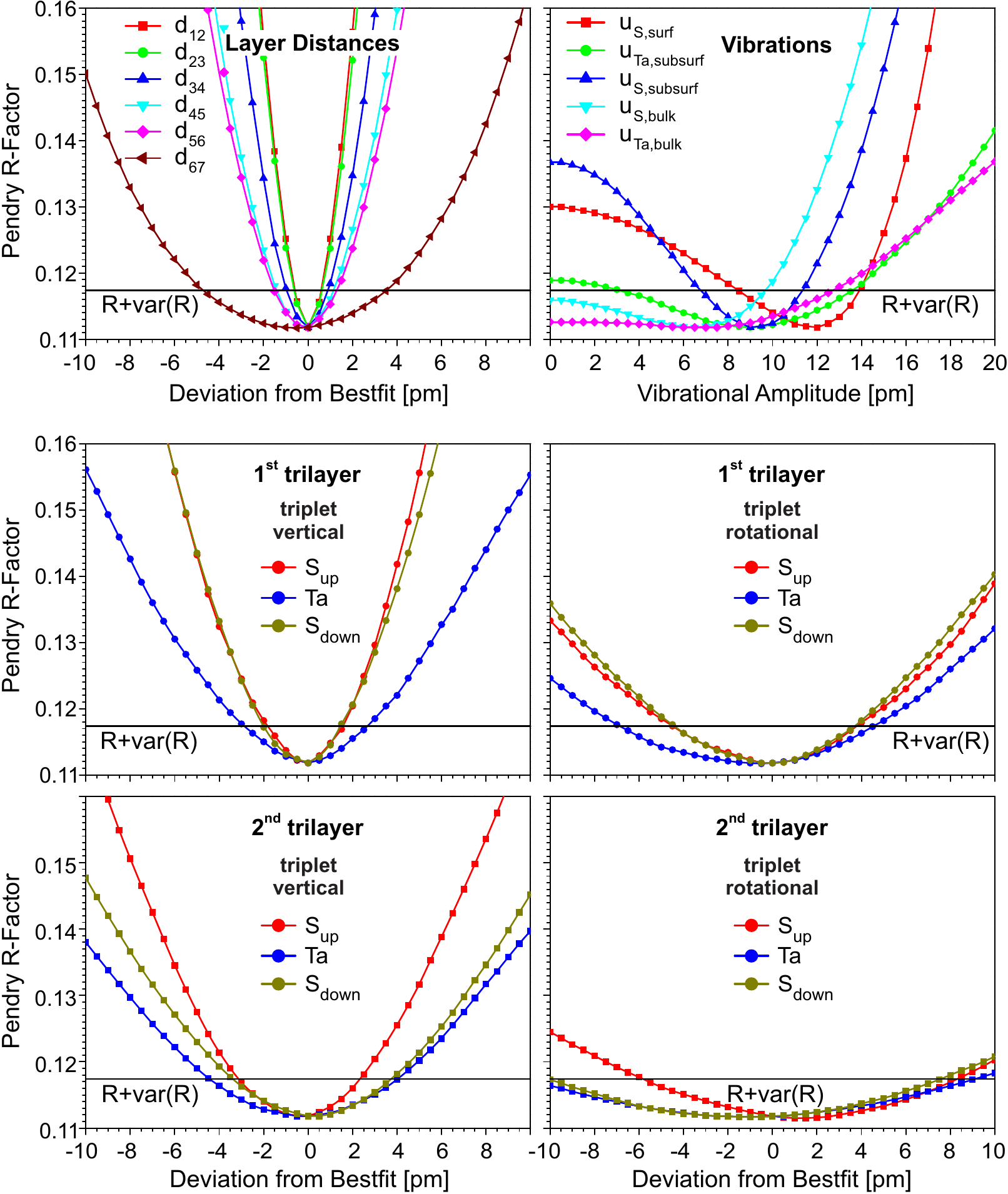}
\caption{(Color online) Top row: Error curves for layer distances
(left) and vibrational amplitudes (right). Below: Error curves for
vertical (left) and rotational lateral displacements (right) of
exemplary triples of atoms (S4, Ta2b) within the first (middle row)
and second trilayer (bottom row).}\label{Errors}
\end{centering}
\end{figure}

Fig.\,\ref{Errors} displays  a selection of so-called ``error
curves'', i.e., the variation of the Pendry R-factor as a function of
certain structural parameters like layer distances (top left),
vibrational amplitudes (top right) or local displacements of triples
of symmetry-related atoms (middle and lower row). A full set of
error curves for all 78 fitted structural parameters is provided in
the Supplemental Material \cite{SupMat}. The parameter values, at which
the error curves intersect the reliability level $R+var(R)$ then
represent the related limits of error. As expected, the sensitivity of
the analysis and thus the width of the error bars scales with their
depth below the surface (due to attenuation) and inversely with
the number of atoms displaced simultaneously. So, layer distances,
i.e., the distance of centers of mass of adjacent S or Ta layers,
each consisting of 13 atoms, have very small errors. They are below
0.01\,\AA\ for the outermost three layer distances, about
0.015\,\AA\ for the next two and only then increase to about
0.04\,\AA\ below the second trilayer (d$_{67}$). For vertical
displacements of only triples of atoms, which are coupled by the
assumed 3-fold rotational symmetry of the trilayers, the error bars are larger, of
course. For the first trilayer, they amount to 0.02\,\AA\ and
0.03\,\AA\ for S and Ta atoms, respectively, and increase to values
of about 0.03\,\AA\ and 0.045\,\AA\ for the second trilayer.

As a general behavior in LEED-IV analyses (performed at normal incidence), 
the sensitivity towards lateral displacements is reduced 
by a factor of 2 - 3 compared to vertical ones. 
This holds also in the present case, where, depending on depth, 
error margins in the range 0.04\,\AA\ - 0.10\,\AA\ result.

The size of the error bar only weakly depends on the specific triple
of atoms as can be seen by inspection of the Supplemental
Material \cite{SupMat}. There is also hardly a difference between
radial or rotational lateral displacements. In this sense, the error
curves displayed in Fig.\,\ref{Errors} can be taken as prototypical.
Only the singulets, i.e., the atoms assuming high symmetry positions
within each layer (Ta1 and S5), have 50-100\,\% larger error bars.

Finally, we see in Fig.\,\ref{Errors} (top right) that the
sensitivity towards vibrational motions is rather low in
particular for bulk atoms. Only the outermost S layer exhibits a
significantly enlarged vibrational amplitude of 0.12\,\AA, while the
subsurface Ta and S atoms of the first trilayer show a minimum for
$u = 0.09$\,\AA. The fit of bulk vibration amplitudes leads to
numerically even lower values of 0.065\,\AA\ with, however, very
little significance, so that the use of common amplitudes for all
atoms below the outermost layer would have only spurious effects on
the fit quality.

\subsection{Local atomic displacements} \label{Structure}

\begin{figure}[tbh]
\begin{centering}
\includegraphics[width=0.30\textwidth]{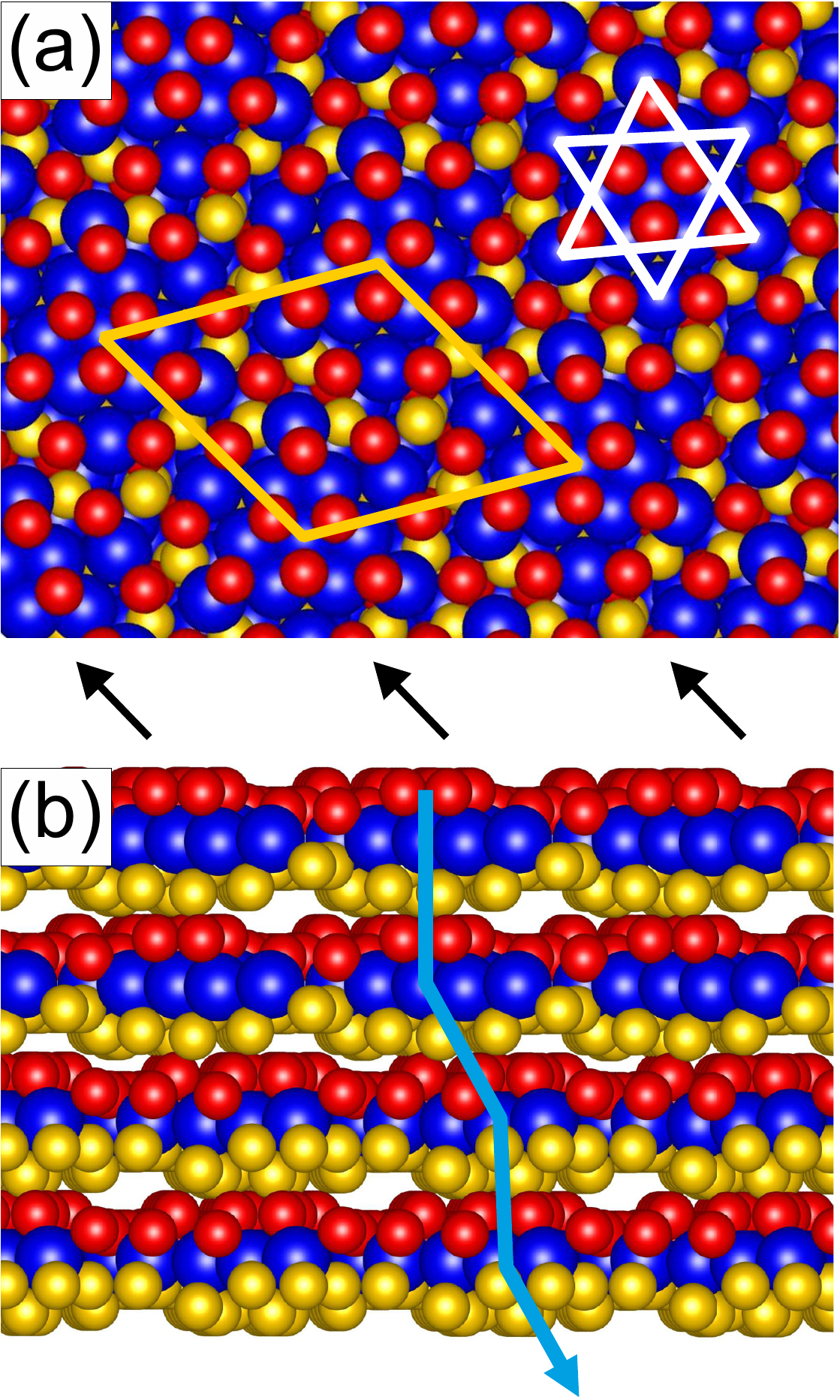}
\caption{(Color online)  (a) Ball model of the bestfit structure in
top view. The atomic displacements from the unreconstructed
positions are five times enlarged for better visibility. Also shown
are the ($\sqrt{13} \times \sqrt{13}$)R13.9$^\circ$ unit mesh
(yellow) and the star-of-David formation of Ta atoms (white). (b)
Side view in direction of a superstructure unit vector (indicated by
black arrows above) and the stacking sequence of layers
(blue).}\label{BestfitModel}
\end{centering}
\end{figure}

\begin{figure}[tbh]
\begin{centering}
\includegraphics[width=0.45\textwidth]{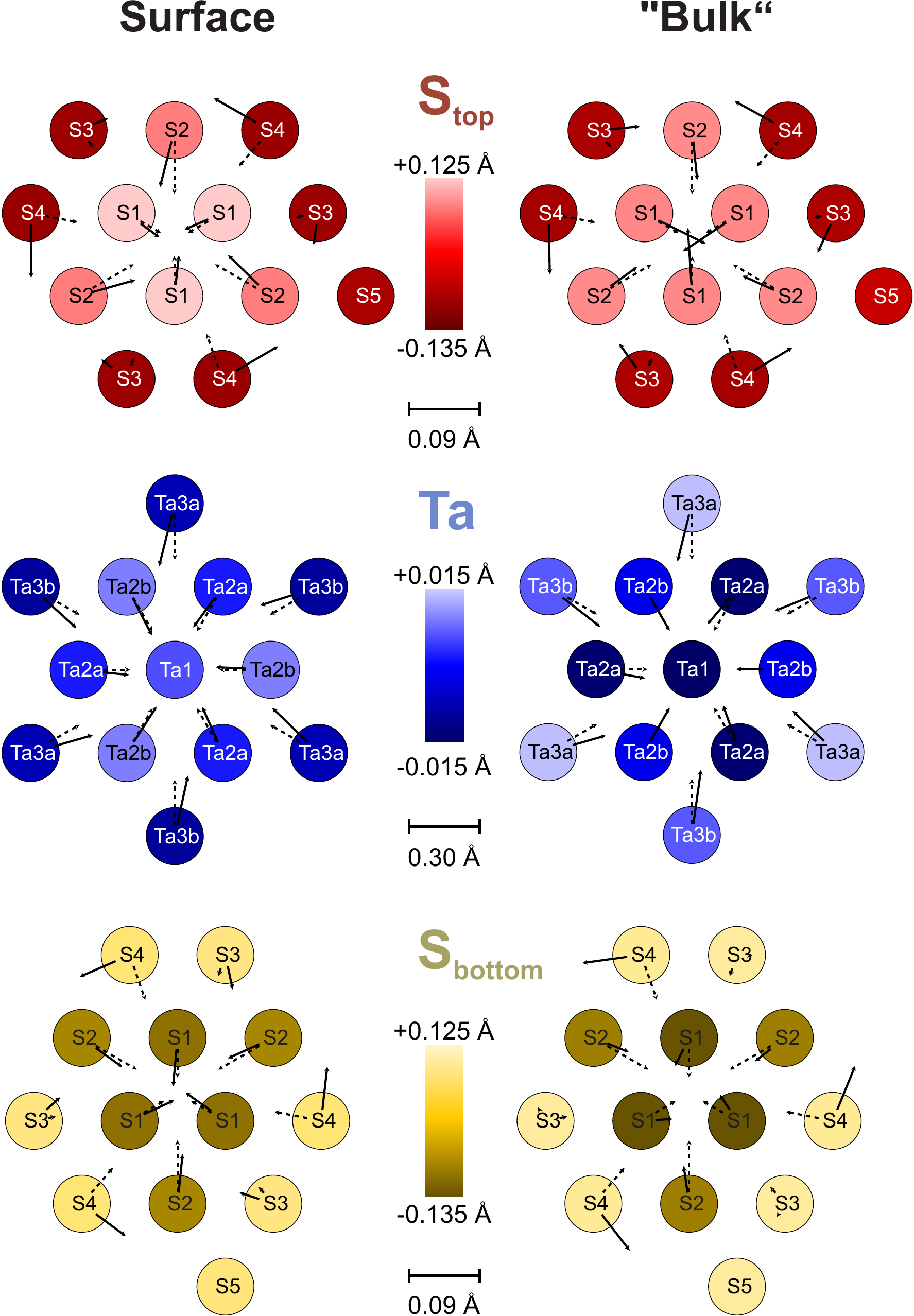}
\caption{(Color online) Schematic representation of the CDW-induced
layer reconstruction within surface (left) and subsurface (``bulk'')
trilayers (right) as derived from the LEED-IV analysis. Vertical
displacements are color-coded, size and direction of the lateral
ones are indicated by solid arrows (lengths according to the scale
bars). Dashed arrows correspond to the XRD analysis of Brouwer
and Jellinek \cite{Brouwer1980}.}\label{Layers}
\end{centering}
\end{figure}

Having fixed the superstructure stacking sequence already in 
Subsection~\ref{Bestfit}, we now want to discuss the bestfit atomic
postions within the two outermost trilayers. A rough impression of the
scenario can be gained from the ball model displayed in
Fig.\,\ref{BestfitModel} in top and side view with local CDW-induced
displacements magnified by a factor of five. For a more quantitative
comparison, the single atomic displacements are visualized in the
schematic drawing of Fig.\,\ref{Layers} by arrows (lateral) and
color-coding (vertical) for every single layer with the respective
scalings given in the center. Moreover, the numerical values are all
tabulated in Tab.\,\ref{Deviations} together with bulk data from
literature \cite{Brouwer1980} for comparison.

As expected, we find every 13 Ta atoms of the unit cell clustered in
a star-of-David-like formation with mostly radial displacements of
the order of 0.2\,\AA\ from their unreconstructed positions
(Tab.\,\ref{Deviations} and Fig.\,\ref{Layers} center row). 
There are only comparatively small positional
differences found for the two Ta layers, though they were completely
independently fitted. Also the correspondence to published data for
the bulk structure \cite{Brouwer1980} (dashed arrows in Fig.\,\ref{Layers}) is very close.

As a consequence of the lateral Ta atom clustering, the S atoms bound within
the contraction zones (S1, S2) are vertically pushed outwards, and vice versa,
they are pulled inwards in the more dilute areas near the edges of
the WS cell (S3 - S5), see Figs.\,\ref{BestfitModel} and
\ref{Layers}. The S atoms also try to follow the Ta atoms laterally,
in order to maintain the high symmetric binding configuration. Such
a movement, however, is largely restricted ($\leq 0.08$\,\AA, note
the different scaling of the arrows for S and Ta atoms in
Fig.\,\ref{Layers}) by the mutual repulsion of S atoms, which are
already rather close packed in the unreconstructed structure.
Consequently, the formerly threefold coordination of S atoms towards
Ta gets lost with variations in the S-Ta bond lengths by up to
0.2\,\AA\ (in particular for S3 and S4; see also
Fig.\,\ref{BestfitModel}(a)).

From the side view in Fig.\,\ref{BestfitModel}(b), we also see clearly
that there cannot be a strict inversion symmetry between top and
bottom S layers (as assumed by Brouwer and
Jellinek \cite{Brouwer1980}) due to the alternate stacking sequence,
which couples the height modulation of S atoms across the
Van-der-Waals gap either in-phase or (close to) anti-phase.
Instead, inversion symmetry should apply for pairs of
trilayers, at least in the bulk. As a consequence, slightly
alternating layer distances within and between trilayers have to be
expected already for the bulk. For the outermost trilayer, this
asymmetry might be further enhanced by surface bond relaxations
caused by the break of translational symmetry. Indeed,
Tab.\,\ref{Deviations} reveals slightly different layer distances
with the upper S layer always somewhat farther from the center Ta
layer than the lower one. While for the second and thus more
bulk-like trilayer this difference (0.016\,\AA) is within the mutual
limits of error, the 0.035\,\AA\ expansion on the outermost S layer
compared to $d_{23}$ appears highly significant (error range $\pm
0.006$\,\AA, c.f. Fig\,\ref{Errors} top left). Also an indication
for the postulated up--down asymmetry of trilayers is the buckling
pattern found for the Ta layers displayed in the center row of
Fig.~\,\ref{Layers}. As expected, it is approximately inverse for the
two adjacent Ta layers. However, their whole corrugation is
too small to be beyond the error margins - a neglection raises the
R-factor in both cases by not more than about 0.002 (a third of the
R-factor's variance).

Another interesting finding of our analysis is that the outermost
trilayer distance (defined as the mutual distance of the corresponding Ta layers: $c_1 = d_{23}+d_{34}+d_{45} = 5.901$\,\AA) is
smaller by almost 0.02\,\AA\ than the bulk trilayer distance $c_0 =
5.919$\,\AA\ of the C-phase \cite{Givens1977}. Remarkably, this value is very
close to the trilayer distance determined for the NC-domain wall
structure ($c_0 = 5.896$\,\AA) \cite{Givens1977}.

\begin{table}[t!]
\centering
\begin{tabular}{c@{\hspace{4mm}}l@{\hspace{4mm}}r@{\hspace{4mm}}r@{\hspace{8mm}}r}
\hline \hline \multicolumn{5}{c} {\textbf{This work}}\\
\hline \hline Layer  & Atom & $d_{rad}$ [\AA] & $d_{rot}$ [\AA] &
$d_\perp$ [\AA] \\ \hline
\multicolumn{5}{l}{~$1^{st}$ trilayer} \\
\hline
1 & S1 & 0.045 & 0.005 & 0.124\\
 & S2 & 0.060 & 0.015 & 0.076 \\
 & S3 & 0.010 & -0.030 & -0.086 \\
 & S4 & 0.015 & 0.075 & -0.089 \\
 & S5 & 0 & 0 & -0.076 \\ \hline
 & & \multicolumn{3}{l}{layer distance $d_{12} = 1.549$\,\AA} \\ \hline
2 & Ta1 & 0 & 0 & 0.006 \\
 & Ta2a & 0.199 & 0.020 & 0.002 \\
 & Ta2b & 0.219 & 0.010 & 0.009 \\
 & Ta3a & 0.245 & 0.061 & -0.008 \\
 & Ta3b & 0.235 & 0.051 & -0.005 \\ \hline
 & & \multicolumn{3}{l}{layer distance $d_{23} = 1.514$\,\AA} \\ \hline
3 & S1 & 0.055 & 0.005 & -0.100 \\
 & S2 & 0.050 & 0.005 & -0.078 \\
 & S3 & 0.025 & -0.015 & 0.080 \\
 & S4 & 0.005 & 0.065 & 0.073 \\
 & S5 & 0 & 0 & 0.075 \\ \hline
 & & \multicolumn{3}{l}{layer distance $d_{34} = 2.850$\,\AA} \\ \hline
\multicolumn{5}{l}{~$2^{nd}$ trilayer} \\ \hline
4 & S1 & 0.080 & -0.005 & 0.083 \\
 & S2 & 0.050 & -0.005 & 0.085 \\
 & S3 & 0.025 & -0.035 & -0.072 \\
 & S4 & 0.015 & 0.070 & -0.080 \\
 & S5 & 0 & 0 & -0.050 \\ \hline
 & & \multicolumn{3}{l}{layer distance $d_{45} = 1.537$\,\AA} \\ \hline
5 & Ta1 & 0 & 0 & -0.015 \\
 & Ta2a & 0.194 & 0.036 & -0.014 \\
 & Ta2b & 0.204 & 0.000 & 0.002 \\
 & Ta3a & 0.225 & 0.056 & 0.014 \\
 & Ta3b & 0.255 & 0.035 & 0.006 \\ \hline
 & & \multicolumn{3}{l}{layer distance $d_{56} = 1.521$\,\AA} \\ \hline
6 & S1 & 0.030 & 0.015 & -0.134 \\
 & S2 & 0.035 & -0.005 & -0.086 \\
 & S3 & -0.005 & -0.015 & 0.096 \\
 & S4 & -0.015 & 0.075 & 0.091 \\
 & S5 & 0 & 0 & 0.096 \\ \hline
 & & \multicolumn{3}{l}{layer distance $d_{67} = 2.839$\,\AA} \\
\hline\hline \multicolumn{5}{c}{\textbf{Brouwer and
Jellinek} \cite{Brouwer1980}}
\\ \hline \hline
 & Atom & \multicolumn{2}{c}{$d_\parallel$ [\AA]~~~} & $d_\perp$ [\AA] \\
\hline
 & Ta1 & \multicolumn{2}{r}{0 \qquad\qquad\qquad} & 0 \\
 & Ta2 & \multicolumn{2}{r}{0.215 \qquad\qquad\qquad} & -0.009 \\
 & Ta3 & \multicolumn{2}{r}{0.236 \qquad\qquad\qquad} & 0.009 \\ \hline
 & S1 & \multicolumn{2}{r}{0.05 \qquad\qquad\qquad} & 0.10 \\
 & S2 & \multicolumn{2}{r}{0.08 \qquad\qquad\qquad} & 0.05 \\
 & S3 & \multicolumn{2}{r}{0.01 \qquad\qquad\qquad} & -0.10 \\
 & S4 & \multicolumn{2}{r}{0.06 \qquad\qquad\qquad} & -0.11 \\
 & S5 & \multicolumn{2}{r}{0 \qquad\qquad\qquad} & -0.07 \\ \hline
\hline \hline
\end{tabular}
\caption{Atomic displacements from the unreconstructed 1\textit{T}-TaS$_2$
structure determined by LEED-IV. Lateral movements ($d_\parallel$)
are divided into radial ($d_{rad}$) and rotational components
($d_{rot}$) w.r.t. the center Ta atom (Ta1) of each trilayer. Layer
distances and vertical shifts are given w.r.t. the single layer's
centers of mass with positive sign directing towards the vacuum. For
comparison the results of Brouwer and
Jellinek \cite{Brouwer1980} for the bulk structure are added below.}
\label{Deviations}
\end{table}

Regarding the lateral displacements, a detailed analysis suffers from
the much larger error margins. Therefore, it is advisable to compare
mainly the overall patterns displayed in Fig.\,\ref{Layers} rather
than single atomic coordinates from Tab.\,\ref{Deviations}. For both
Ta and S layers, we see quite similar patterns in each case with
mutual deviations much smaller than the single atomic error margins,
which further increases the confidence into the numerical results of
this analysis.

For Ta atoms the lateral displacements are mainly radial and caused
by the local contraction. However, superimposed there is also an
unidirectional azimuthal movement shifting the outer Ta3 atoms with
0.04\,\AA\ - 0.06\,\AA\ about twice as far as the inner Ta2 atoms.
In total, this describes a small but concerted rotation of the whole
Ta cluster within the `$\sqrt{13}$\,'-supercell. For S atoms, there
are even somewhat larger azimuthal displacements though with
varying directions. In particular, the sulphur triplet S4 is found to
move in all layers almost exclusively rotational with amplitudes of
about 0.07\,\AA. Most probably, the common rotation is overlaid by
local forces caused by the asymmetric binding configurations of S
atoms mentioned above. At least the largest rotational displacements
of Ta and S atoms within the 1$^{st}$ trilayer are well beyond the
error margins, and even for the 2$^{nd}$ trilayer they are still clearly detectable though just within the overall error limit (cf. Fig.~\,\ref{Errors} center and bottom right). Thus, the rotation of the clusters is indeed a
statistically significant property of the CDW-induced structure.

\section{Conclusions}

The full-dynamical LEED-IV analysis presented here provides the
first quantitative analysis for the surface and to some extent also
for the bulk structure of the low-temperature commensurate CDW phase of
1$T$-TaS$_2$. Due to the enormous data basis used for the analysis and
the high fit quality achieved, we can determine all atomic positions 
within the near-surface regime, i.e., for the outermost two trilayers, 
with the precision of a few picometers. Our analysis reveals that the
well-accepted star-of-David-like reconstruction of Ta atoms also
holds for the surface with some small surface-specific
modifications. In particular, we prove statistically significant 
an expansion of the outermost S layer and a reduced 1$^{st}$-2$^{nd}$ 
trilayer distance. We further find clear evidence that the CDW not only
induces a radial contraction of every 13 Ta atoms towards clusters in
a ($\sqrt{13} \times \sqrt{13}$)R13.9$^\circ$ periodic arrangement,
but also causes a certain common rotation of these clusters within
the cell.

The most important and clearest finding of the analysis is that the
surface obviously pins a vertical stacking of the CDW-induced
`$\sqrt{13}$\,'-supercells between the 1$^{st}$ and 2$^{nd}$
trilayer. The stacking between 2$^{nd}$ and 3$^{rd}$ trilayer, in
contrast, turned out to be much more laborious to elucidate and
could be resolved with statistical significance only because of the
excellent fit quality. It involves a lateral shift of almost half a
superstructure unit vector towards one of the equivalent registry
positions 7, 8, or 11 according to the labeling of
Fig.\,\ref{Nomenclature}(d). The associated break of 3-fold symmetry
is located more than 1\,nm  below the surface and thus leads to only
tiny intensity deviations of symmetry-related spots in the LEED
pattern, which are below the detection limit of the present
analysis. So, we cannot differentiate here between 
a real mono-domain termination or, alternatively, a domain mixture
of extended patches with all three (subsurface) stacking
orientations of the \{7,8,11\} class. 

The presented results have two important implications 
on the surface electronic structure of the low-temperature 
commensurate CDW phase of 1$T$-TaS$_2$.
First, it is a remarkable finding that a vertically on-top-stacked double-trilayer 
is pinned at the surface as this structural unit has been identified 
as the driver of CDW stacking-induced energy-gap formation 
at the Fermi level \cite{Ritschel2015,Ritschel2018,Lee2019}.
Regarding the question about the nature of the ground state 
of bulk 1$T$-TaS$_2$, which density functional theory calculations predicts 
to be a Peierls-type insulator \cite{Darancet2014,Ritschel2018,Lee2019}, 
whereas (surface-sensitive) time- and angle-resolved photoemission spectroscopy results 
indicate dominance of local Mott physics \cite{Perfetti2006,Petersen2011,Hellmann2012,Ligges2018}, 
this appears to favor the former interpretation.
However, it is also possible that the atomic surface structure 
reported here gives rise to a surface Mott phase. 
Second, the obtained surface structure clearly reveals 
a local inversion asymmetry in the surface trilayer 
due to an asymmetry in the Ta--S layer separations.
In principle, the relatively strong spin-orbit coupling in the material 
can then induce a momentum-dependent spin splitting 
resulting in spin-polarized surface states \cite{Riley2014}. 
This calls for spin- and angle-resolved photoemission spectroscopy measurements.
The detailed atomic surface structure thus emerges as an important ingredient 
in any understanding of the intriguing electronic phenomena probed by surface-sensitive techniques  \cite{Cho2016,Ma2016,Gerasimenko2019,Perfetti2006,Petersen2011,Hellmann2012,Ligges2018,Vogelgesang2018,Ritschel2018,Ngankeu2017}.
There is clearly a need for further theoretical and experimental scrutiny 
of the atomic and electronic surface structure of 1$T$-TaS$_2$ and related materials, 
both for bulk and few-trilayer crystals.

\vspace{5mm}

This work was supported by the European Research Council (ERC StG, ID: 639119).

\appendix
\section{Experimental details} \label{ExpDetails}

The TaS$_2$ sample (diameter about 3\,mm, thickness 0.5\,mm) was grown 
as described in the supplement of Ref.\,\onlinecite{Ligges2018}. Here, it was
mounted on a copper support plate at the sample transfer holder
using an electrically conducting two-component adhesive. 
At the top of the sample a ceramic stick was
fixed with the same adhesive. The transfer holder was then
introduced into the UHV ($p < 2\cdot10^{-10}$\,mbar) chamber via a
two-stage load-lock system and mounted on the manipulator, which
allowed for high-precision x-, y-, z-translations, rotation around
an axis within the sample's surface and tilt against its normal. The
sample was cooled by direct contact to a liquid nitrogen
reservoir and heated either by a hot filament or by additional
electron bombardment from the rear, whereby the temperature was
monitored using a K-type thermocouple attached to the Cu base plate. 
After careful degassing of the sample holder, the TaS$_2$ crystal was cooled down to about 100\,K and subsequently cleaved by hitting the ceramic stick glued on the crystal with a
wobble-stick. In order  to avoid any time loss prior to the LEED experiment the cleavage was performed right in front of the LEED optics.

\begin{figure}[htb]
	\begin{centering}
		\includegraphics[width=0.4\textwidth]{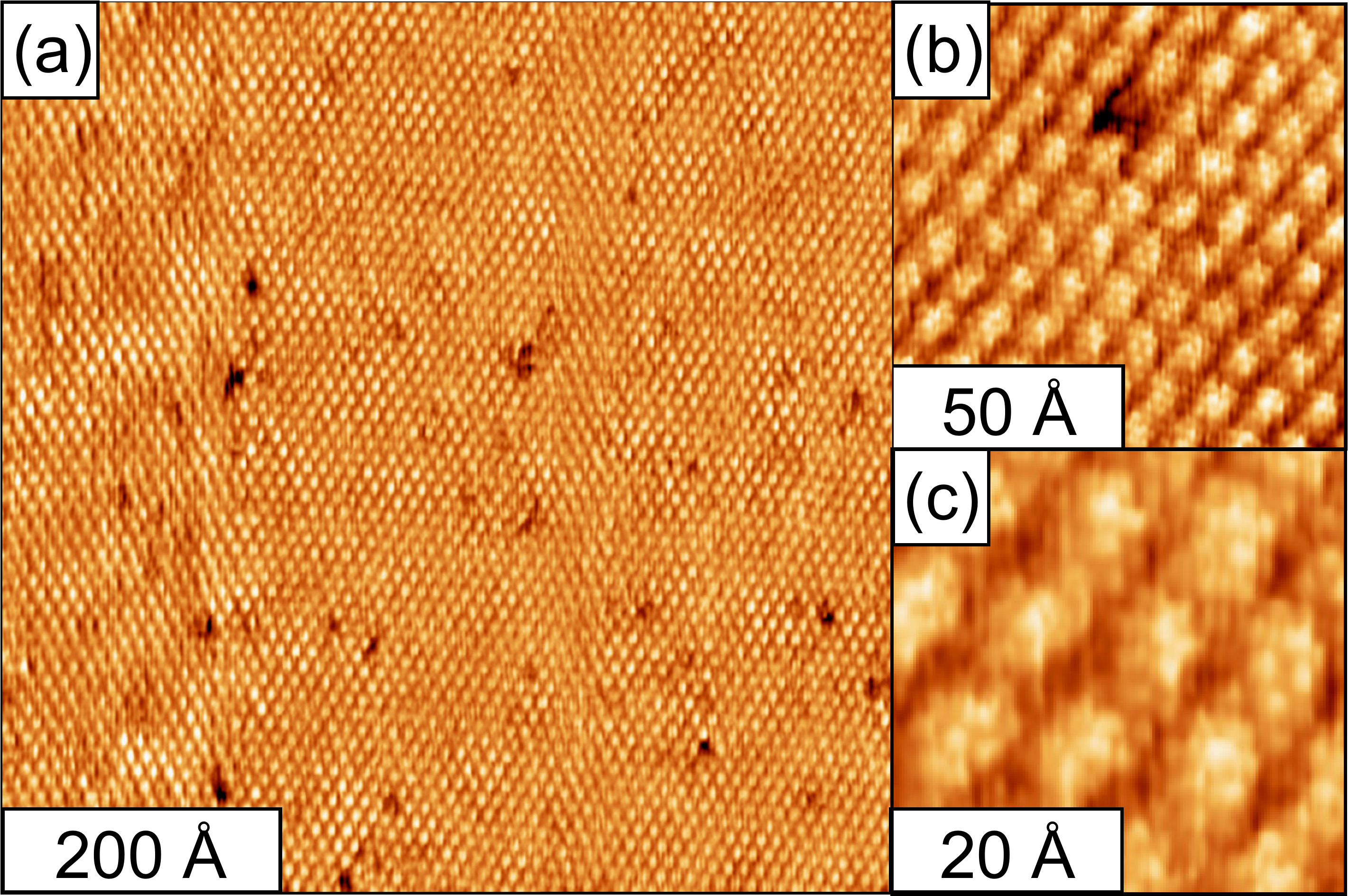}
		\caption{(Color online) (a) Room-temperature STM survey image of the
			TaS$_2$ sample taken 30\,h after cleavage (U$_{sample}$ = -90\,mV; I
			= 1.3\,nA). (b,c) Same as (a) but with atomic resolution.
			(U$_{sample}$ = +5\,mV; I = 3.4\,nA).}\label{STM}
	\end{centering}
\end{figure}

After completion of all LEED experiments described in Subsection~\ref{Experiment} 
($\approx 1$ day after cleavage) the sample was transferred  to a room-temperature STM 
within the same UHV apparatus for further characterization 
of the surface morphology. Unfortunately, at room temperature 
the C-phase is no more stable, so that only the NC-phase could be imaged here. 
Large-scale STM topographs (Fig.~\ref{STM}(a)) 
were taken at different places of the sample. These always showed the
pseudo-regular arrangement of `$\sqrt{13}$\,'-unit cells with
spatially varying contrast typical for the long-range domain
structure of the NC-phase \cite{Burk1991,Burk1994,Bando1997}. The
moir\'{e}-type atomic configuration of topmost sulphur atoms can
directly be seen in the atomically resolved picture displayed in
Fig.~\ref{STM}(b). STM also reveals a small density of randomly
distributed defects imaged as depressions with an apparent depth of
about 2\,\AA\ and typical sizes of the order of whole
`$\sqrt{13}$\,'-unit cells. Since clusters of missing sulphur atoms
seem rather improbable, we tentatively attribute these depressions
to spurious contaminations adsorbed during the one day exposure to
the residual gas atmosphere of the UHV chamber. Remarkably, 
we did not find any single atomic step during the entire STM investigation 
performed at various places across the crystal's surface.

\section{Computational details} \label{CompDetails}
Phase shifts  were calculated by J. Rundgren's program
\emph{EEASiSSS} \cite{Rundgren} for an unreconstructed 
1\textit{T}-TaS$_2$ crystal. Due to the comparably low maximum energy of 250\,eV, 
values of $ \ell_{\text{max}}$~=~9 turned out to be sufficient. 
This program also calculates self-consistently 
the corresponding dependence of the real part of
the inner potential on the electron energy E: $V_\text{0r} =
max[(-0.05-63.42/\sqrt{E+17.15}, -9.00)$\,eV. To account for the
filament's work function and calibration uncertainties a constant
offset $V_\text{00}$ for the electron energy was adjusted 
within the analysis resulting in a value of
$V_\text{00} = -3.0$\,eV. The damping of the electrons due to
inelastic processes was considered by a constant optical potential
$V_\text{0i}$, which was fitted to 4.75\,eV in a later stage of the
analysis. As lateral lattice parameter we used
a$_{\text{TaS$_2$}}=$~3.354\,\AA, which was extrapolated from the
temperature dependence given in Ref.~\onlinecite{Givens1977} to our data
acquisition temperature of 100\,K. A vertical lattice parameter $c$
was not introduced to the calculations but resulted automatically
from the fit of individual layer distances.

 For better convergence of the fit, vertical, parallel, and rotational
displacements were optimized separately in an iterative process,
whereby the grid widths were successively reduced down to
0.0025\,{\AA} for vertical and 0.005\,{\AA} for lateral parameters,
since for the latter, normal incidence LEED is less sensitive. The
iterations converged rapidly, indicating that parameter dependencies
between the different parameter classes are quite small.
Additionally, the (common) vibrational amplitude of the surface S
atoms was adjusted, while the vibrations of bulk atoms were first
kept fixed at values derived from Ref.~\onlinecite{Spijkerman1997}.
Finally, and only for the bestfit configuration, the vibrational
amplitudes (taken as isotropic) of all three layers of the topmost trilayer as well as
for bulk Ta and S atoms were optimized without readjusting the
geometrical parameters. Also, the angular spread within the
slightly convergent primary beam used in the LEED experiment was
accounted for by a small off-normal angle of incidence (here fitted
to 0.51$^\circ$) and averaging over 6 azimuthal directions of
incidence ($\Phi = 0^\circ, 60^\circ$, ...).

The parameter variations for error determination were performed in the
TensorLEED approximation and as usual with all other parameters held
fixed at their bestfit value. This procedure saves enormous
computational time, but as a drawback it neglects possible parameter
correlations. Experience tells that such correlations increasingly
diminish with the size of the data base used for the analysis, which
is with more than 15\,keV quite huge in our case. Moreover, in the
course of fitting we did not find any hint for significant parameter
coupling, so that we have confidence in our error
estimate. We also note that, again for sake of
computational time minimization, the R-factor calculation was not
performed for the azimuthally averaged spectra but for one single
azimuth of incidence ($\Phi = 0^\circ$) only. For this setup the minimum
R-factor was slightly higher ($R = 0.112$) and so also the
reliability level $R+var(R) = 0.118$. This, however, means in
effect just a practically rigid upward shift of the curves by
$\Delta R = 0.002$ with no measurable consequences for the error bar
determination.

\section{Trilayer symmetry and CDW stacking} \label{Stacking}

A prerequisite for any LEED-IV structure determination is 
a proper analysis of the symmetry of considered structural models, 
in order to identify symmetry-related groupings of atoms and
parameter constraints. It should be noted that the mere existence of
the surface only allows for surface-normal mirror or glide planes as
well as rotation axes.

The unreconstructed 1\textit{T}-TaS$_2$ structure consists of primitive
hexagonal sulphur or tantalum layers. They are packed to S--Ta--S
trilayers with an internal fcc-like A--B--C layer stacking, which
reduces the p6m symmetry of the single layers to p3m. 
With the formation of the CDW-induced ($\sqrt{13} \times
\sqrt{13}$) reconstruction with 13.9$^\circ$ rotational angle any
mirror plane gets lost, independent of the detailed nature of the
distortion. Therefore, the new supercell has at maximum a 3-fold rotational
symmetry (p3). Consequently, two symmetrically equivalent superstructure domains are
possible (both displayed in Fig.~\ref{Nomenclature}(a)) being rotated by
either +13.9$^\circ$ or -13.9$^\circ$ with respect to the unit mesh
of the basic (unreconstructed) lattice.

The Wigner-Seitz (WS) cell of the `$\sqrt{13}$\,'-super\-structure
contains 39 atoms per trilayer - 13 Ta- and 26 S-atoms - as shown in
Fig.~\ref{Nomenclature}(b). Rotation axes intersect the WS cell in
the center and the corners, thus in each single Ta- or S-layer there
is just one atom of the WS cell on a rotation axis.
All other atoms must be grouped into triples linked by
the 3-fold rotational symmetry. In total, there are 5
symmetrically inequivalent atomic sites per single S or Ta layer (15
per trilayer) labeled in Fig.~\ref{Nomenclature}(b). This labeling
of sites is adopted from Brouwer and
	Jellinek \cite{Brouwer1980}, who, however, treated the Ta layer
(inaccurately) as sixfold symmetric. Hence, we use extra
labels ``a'' and ``b'' for discrimination.

In unreconstructed 1\textit{T}-TaS$_2$ the trilayers are stacked 
vertically on top of each other. Any CDW-induced reconstruction 
is not expected to lead to a change of this principal layer stacking, 
since that would require a concerted movement of entire S or Ta layers.
Nevertheless, such stacking faults have occasionally been observed 
by scanning transmission electron microscopy
in exfoliated 1\textit{T}-TaS$_2$ samples \cite{Hovden2016}. Therefore, 
it cannot be excluded \textit{a priori} that such a stacking fault 
might be induced near the very surface by the cleavage process, 
an additional option, which has to be regarded in the fitting process.
A completely different question, however, is the stacking of the
`$\sqrt{13}$\,'-\emph{supercells}, i.e., whether there is a lateral
phase shift between CDWs of adjacent trilayers. In total,
there are 13 different possible registries, which can easily be
visualized by choosing the alternative representation of the
translational unit cell shown in Fig.~\ref{Nomenclature}(c) (light green rhomb). A
back-projection of these registries into the WS cell is displayed in
Fig.~\ref{Nomenclature}(d). Obviously, any registry except the
vertical stacking \{0\} breaks the 3-fold rotational symmetry of the
single trilayer and leaves the surface model without any remaining
symmetry element (p1).

The internal 3-fold rotational symmetry of each trilayer, however,
will hardly be affected by the symmetry break across the
Van-der-Waals gap. Thus, the various stackings can be grouped again
into 5 classes of symmetrically inequivalent configurations, which
are \{0\}, \{1,3,9\}, \{4,10,12\}, \{2,5,6\}, and \{7,8,11\}
according to the labels given in Fig.~\ref{Nomenclature}(d). While
the supercell stacking in the commensurate CDW phase of 1\textit{T}-TaS$_2$
could not be unequivocally resolved by X-ray \cite{Tanda1984} or
electron diffraction studies \cite{Fung1980}, it was revealed by
transmission electron microscopy \cite{Ishiguro1991} as alternating
between the classes \{0\} and \{7,8,11\}. In their study,
Ishiguro and Sato \cite{Ishiguro1991} found that with very few exceptions, 
every vertical stacking is followed by another with lateral shift, 
whereby both alternatives, i.e., ``sliding'' (e.g., 0,8,0,8,0,8,...) as well as
``screw'' stacking (e.g., 0,7,0,8,0,11,...), occurred locally indicating a partially disordered vertical stacking.

\section{Fitting procedure and model discrimination} \label{Fitting}

In order to save some computational time we performed 
the rough model exclusion with a reduced data set consisting of 66 beams only, 
which is about half of the total data base. In a first round, 
we investigated models with different superstructure stackings 
between the first and second trilayer and - for simplicity - 
vertical stacking below. We sequentially adjusted
vertical and radial displacements of atoms within each model,
disregarding any rotational degree of freedom. It turned out that a
vertical superstructure stacking between the outermost two trilayers
resulted in a much lower R-factor value of $R = 0.191$ compared to
values of $R = 0.270 - 0.338$ for the four other stacking classes,
cf. upper part of Table\,\ref{RFactor}. At that stage of fitting, 
the variance of the current best R-factor was already as low as 
$var(R) = 0.013$, i.e., all models with R-factor values way above 
$R + var(R) = 0.204$ can safely be excluded.

\begin{table}[hbt]
	\centering
	\renewcommand{\arraystretch}{1.3}
	\begin{tabular}{l@{\hspace{2mm}}c@{\hspace{2mm}}c@{\hspace{2mm}}c@{\hspace{2mm}}c@{\hspace{2mm}}c}
		\hline \hline \multicolumn{6}{c} {1$^{st}$ - 2$^{nd}$ trilayer stacking}\\
		\hline Stacking class  & \textbf{\{0\}} & \{1,3,9\} & \{4,10,12\} & \{2,5,6\} & \{7,8,11\} \\
		R-factor  & \textbf{0.191} & 0.284 & 0.306 & 0.270 & 0.338\\  \hline
		\hline
		\multicolumn{6}{c} {2$^{nd}$ - 3$^{rd}$ trilayer stacking}\\
		\hline Stacking class  & \{0\} & \{1,3,9\} & \{4,10,12\} & \{2,5,6\} & \textbf{\{7,8,11\}} \\
		R-factor  & 0.160 & 0.166 & 0.154 & 0.143 & \textbf{0.139} \\
		R(fine-grid) & --- & --- & --- & 0.132 & \textbf{0.125} \\
		R(all data) & --- & ---& --- & 0.131 & \textbf{0.123} \\
		R(angular corr.) & --- & ---& --- & --- & \textbf{0.113} \\
		R(vibr. opt.) & --- & ---& --- & --- & \textbf{0.110}\\\hline \hline
	\end{tabular}
	\renewcommand{\arraystretch}{1.0}
	\caption{Pendry R-factor values achieved for different models of
		superstructure stacking and stages of fit refining. For details see
		text.} \label{RFactor}
\end{table}

Next, we compared all different stacking models for the 2$^{nd}$ to
3$^{rd}$ trilayer positioning while keeping the superstructure
stacking of the outermost trilayer fixed at the vertical class
\{0\}. Note that all these models have
the outermost six S or Ta layers in common, i.e., they start to
differ from each other only at a depth of about 12\,\AA, where the
electron wave-field has already been greatly attenuated by inelastic
scattering. Hence, the R-factor differences between different models 
will be much smaller. Therefore, the fit was performed at a
more elaborate level by additionally regarding rotational atomic
movements with respect to the central Ta atom. This reduced the
overall R-factor level, e.g., $R = 0.191 \rightarrow 0.160$ for the
``all-vertical'' model. As can be seen in the lower part of
Table\,\ref{RFactor}, there are still significant differences in the
R-factor values for the various models. The bestfit ($R = 0.139$) 
is achieved for a \{7,8,11\} stacking. However, 
also a \{2,5,6\} stacking would be consistent with the data, 
given the variance of the R-factor at this fit level ($var(R) =
0.0095$). All other stackings can already be excluded.

For these two remaining models, we then performed the structural
search on a finer grid (0.005\,\AA\ for lateral and 0.0025\,\AA\ for
vertical displacements) around the respective former bestfit
configurations, which brought both R-factor values further down and
also increased their relative distance, though still not beyond the
variance level. This, however, could be reduced by
increasing the data basis of the fit. With the whole available 
data of more than 15\,keV in total, the models finally differed 
by $\Delta R = 0.008$, while the variance, scaling inversely 
with the square root of the data base size, became as low as $var(R) = 0.006$.

Although the R-factor variance level is by no means a sharp limit
for the unambiguous exclusion of a model we have strong confidence
that the \{7,8,11\} class is indeed the correct description for the
2$^{nd}$ to 3$^{rd}$ trilayer stacking. This is backed by the observation 
that the bestfit for the \{2,5,6\} model produced physically rather
implausible atomic displacements in parts, which did not occur for the
\{7,8,11\} class. Finally, it is also quite striking that we just
find the very same stacking sequence at the surface as
known for the bulk \cite{Ishiguro1991}, cf. Appendix~\ref{Stacking}.

For further reducing the R-factor value of the bestfit model - and
with that the error margins of the analysis - we eventually
corrected the intensity calculations for effects of the slightly
convergent beam of incoming electrons and optimized the vibrational
amplitudes of surface and bulk S and Ta atoms.

\clearpage
\onecolumngrid

\begin{center}
\textbf{SUPPLEMENTARY INFORMATION}
\end{center}
\noindent This supplement comprises tabulated coordinates of the the bestfit
structure as well as plots of all experimental and calculated
bestfit spectra and ``error curves'' for all fitted atomic
displacements.

\clearpage
\section*{Atomic coordinates of the LEED bestfit structure}

\begin{minipage}[t]{0.5\textwidth}
\renewcommand{\arraystretch}{1.25}
\begin{center}
\begin{tabular}{l@{\hspace{5mm}}r@{\hspace{5mm}}r@{\hspace{5mm}}r}
\hline
{Atom}  & \multicolumn{1}{l}{~X [\AA]}   &  \multicolumn{1}{l}{~Y [\AA]}& \multicolumn{1}{l}{~Z [\AA]}  \\[.5ex]
\hline\hline
S11     &   1.6355      &   0.9500      &   0.0000   \\
S14     &   3.2871      &   3.9100      &   0.2125   \\
S13     &   5.0264      &   0.9370      &   0.2100   \\
S12     &   3.3096      &   -1.8935     &   0.0475   \\
S11     &   0.0050      &   -1.8914     &   0.0000   \\
S11     &   -1.6405     &   0.9415      &   0.0000   \\
S12     &   -0.0150     &   3.8129      &   0.0475   \\
S13     &   1.6355      &   -4.8216     &   0.2100   \\
S14     &   -5.0298     &   0.8918      &   0.2125   \\
S15     &   6.7076      &   -1.9363     &   0.2000   \\
S14     &   1.7425      &   -4.8015     &   0.2125   \\
S12     &   -3.2946     &   -1.9194     &   0.0475   \\
S13     &   -3.3246     &   3.8846      &   0.2100   \\
\hline
Ta21    &   0.0000      &   0.0000      &   1.6675   \\
Ta22a   &   1.5600      &   2.7417      &   1.6715   \\
Ta22b   &   3.1345      &   0.0100      &   1.6640   \\
Ta22a   &   1.5945      &   -2.7218     &   1.6715   \\
Ta22b   &   -1.5586     &   -2.7195     &   1.6640   \\
Ta22a   &   -3.1543     &   -0.0199     &   1.6715   \\
Ta22    &   -1.5760     &   2.7095      &   1.6640   \\
Ta23a   &   -0.0598     &   5.5635      &   1.6810   \\
Ta23b   &   4.8022      &   2.8301      &   1.6785   \\
Ta23a   &   4.8480      &   -2.7299     &   1.6810   \\
Ta23b   &   0.0499      &   -5.5737     &   1.6785   \\
Ta23a   &   -4.7882     &   -2.8336     &   1.6810   \\
Ta23b   &   -4.8521     &   2.7437      &   1.6785   \\
\hline
S31     &   1.6318      &   -0.9364     &   3.2855   \\
S32     &   3.3082      &   1.9157      &   3.2630   \\
S34     &   5.0376      &   -0.9034     &   3.1130   \\
S33     &   3.3261      &   -3.8635     &   3.1055   \\
S32     &   0.0050      &   -3.8229     &   3.2630   \\
S31     &   -1.6268     &   -0.9450     &   3.2855   \\
S31     &   -0.0050     &   1.8814      &   3.2855   \\
S33     &   1.6828      &   4.8124      &   3.1055   \\
S33     &   -5.0090     &   -0.9488     &   3.1055   \\
S34     &   -1.7364     &   4.8145      &   3.1130   \\
S35     &   1.6769      &   -6.7771     &   3.1105   \\
S34     &   -3.3013     &   -3.9111     &   3.1130   \\
S32     &   -3.3131     &   1.9071      &   3.2630   \\
\hline\hline
\end{tabular}
\end{center}
\end{minipage}
\hfill
\begin{minipage}[t]{0.5\textwidth}
\renewcommand{\arraystretch}{1.25}
\begin{center}
\begin{tabular}{l@{\hspace{5mm}}r@{\hspace{5mm}}r@{\hspace{5mm}}r}
\hline
{Atom}  & \multicolumn{1}{l}{~X [\AA]}   &  \multicolumn{1}{l}{~Y [\AA]}& \multicolumn{1}{l}{~Z [\AA]}  \\[.5ex]
\hline\hline
S41     &   1.6101      &   0.9238      &   5.9525   \\
S44     &   3.2909      &   3.9067      &   6.1150   \\
S43     &   5.0125      &   0.9291      &   6.1075   \\
S42     &   3.3080      &   -1.9156     &   5.9500   \\
S41     &   -0.0050     &   -1.8565     &   5.9525   \\
S41     &   -1.6051     &   0.9325      &   5.9525   \\
S42     &   0.0050      &   3.8228      &   5.9500   \\
S43     &   -1.7016     &   -4.8055     &   6.1075   \\
S44     &   -5.0288     &   0.8968      &   6.1150   \\
S45     &   6.7076      &   -1.9363     &   6.0850   \\
S44     &   1.7379      &   -4.8034     &   6.1150   \\
S42     &   -3.3129     &   -1.9070     &   5.9500   \\
S43     &   -3.3110     &   3.8763      &   6.1075   \\
\hline
Ta51    &   0.0000      &   0.0000      &   7.5875   \\
Ta52a   &   1.5495      &   2.7533      &   7.5865   \\
Ta52b   &   3.1494      &   0.0000      &   7.5740   \\
Ta52a   &   1.6097      &   -2.7185     &   7.5865   \\
Ta52b   &   -1.5748     &   -2.7274     &   7.5740   \\
Ta52a   &   -3.1591     &   -0.0349     &   7.5865   \\
Ta52b   &   -1.5747     &   2.7275      &   7.5740   \\
Ta53a   &   -0.0548     &   5.5833      &   7.5585   \\
Ta53b   &   4.7923      &   2.8069      &   7.5660   \\
Ta53a   &   4.8626      &   -2.7442     &   7.5585   \\
Ta53b   &   0.0348      &   -5.5536     &   7.5660   \\
Ta53a   &   -4.8078     &   -2.8392     &   7.5585   \\
Ta53b   &   -4.8271     &   2.7467      &   7.5660   \\
\hline
S61     &   1.6584      &   -0.9402     &   9.2275   \\
S62     &   3.3260      &   1.9146      &   9.1800   \\
S64     &   5.0590      &   -0.8975     &   9.0025   \\
S63     &   3.3457      &   -3.8863     &   8.9975   \\
S62     &   -0.0050     &   -3.8378     &   9.1800   \\
S61     &   -1.6434     &   -0.9661     &   1.6355   \\
S61     &   -0.0150     &   1.9063      &   9.2275   \\
S63     &   1.6927      &   4.8405      &   8.9975   \\
S63     &   -5.0383     &   -0.9544     &   8.9975   \\
S64     &   -1.7523     &   4.8300      &   9.0025   \\
S65     &   1.6769      &   -6.7771     &   8.9975   \\
S64     &   -3.3067     &   -3.9325     &   9.0025   \\
S62     &   -3.3211     &   1.9231      &   9.1800   \\

\hline \hline
\end{tabular}
\end{center}
\end{minipage}
\vspace{5mm}

\noindent Coordinates of atoms within the outermost two trilayers as
resulting from the LEED analysis. The z-axis points towards the
crystal's bulk. The third trilayer was taken identical to the second
one, but laterally shifted by $d=6.7076$\,\AA. The vertical distance
between atoms S61 and S72 ($\equiv S42$) was fitted to $2.6195$\,\AA.

\clearpage

\section*{Comparison of experimental and calculated bestfit spectra}

\begin{minipage}[t]{1\textwidth}
\begin{center}
\includegraphics[width=0.85\textwidth]{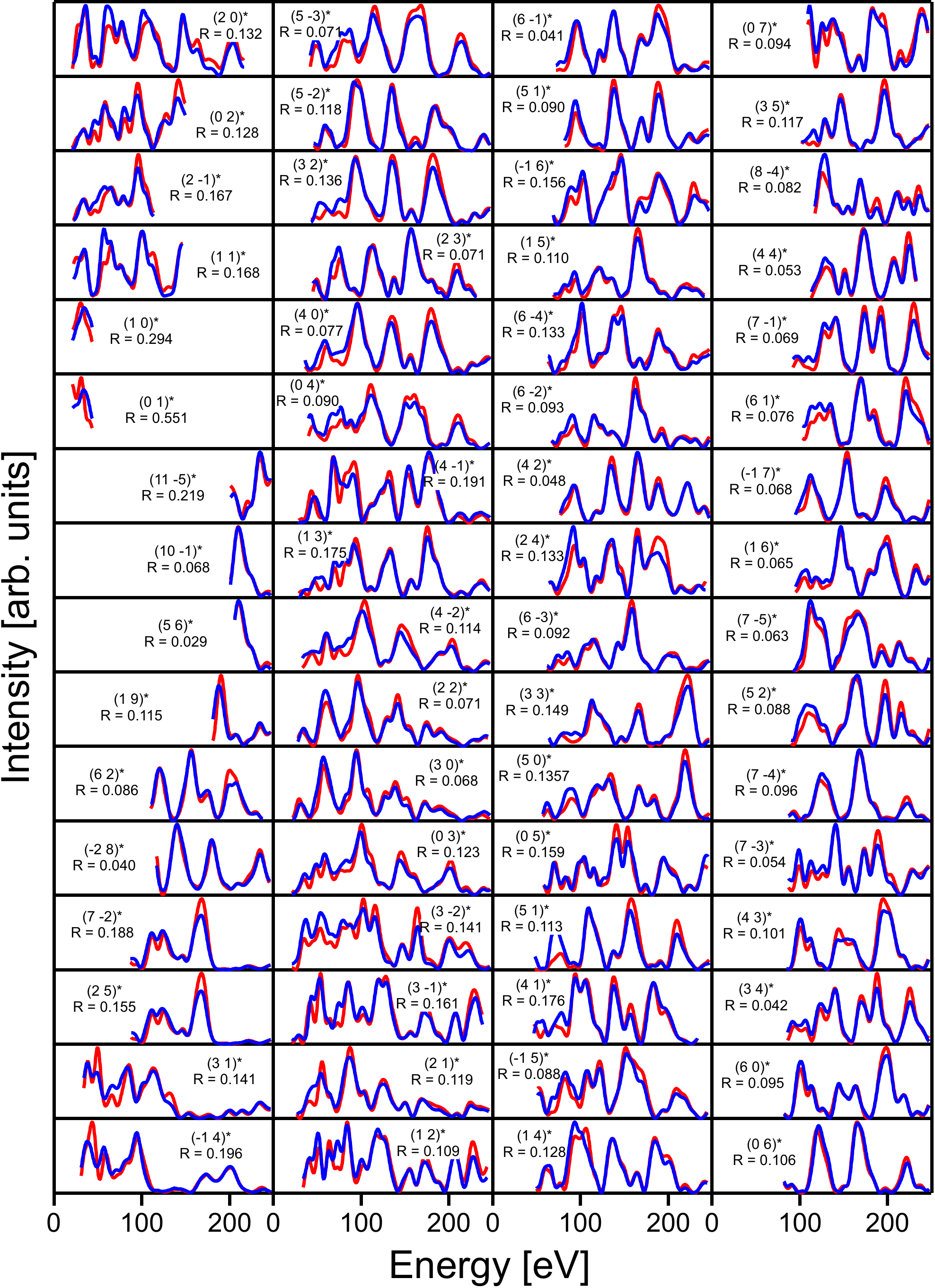}
\end{center}
\end{minipage}

\begin{minipage}[t]{1\textwidth}
\begin{center}
\includegraphics[width=0.85\textwidth]{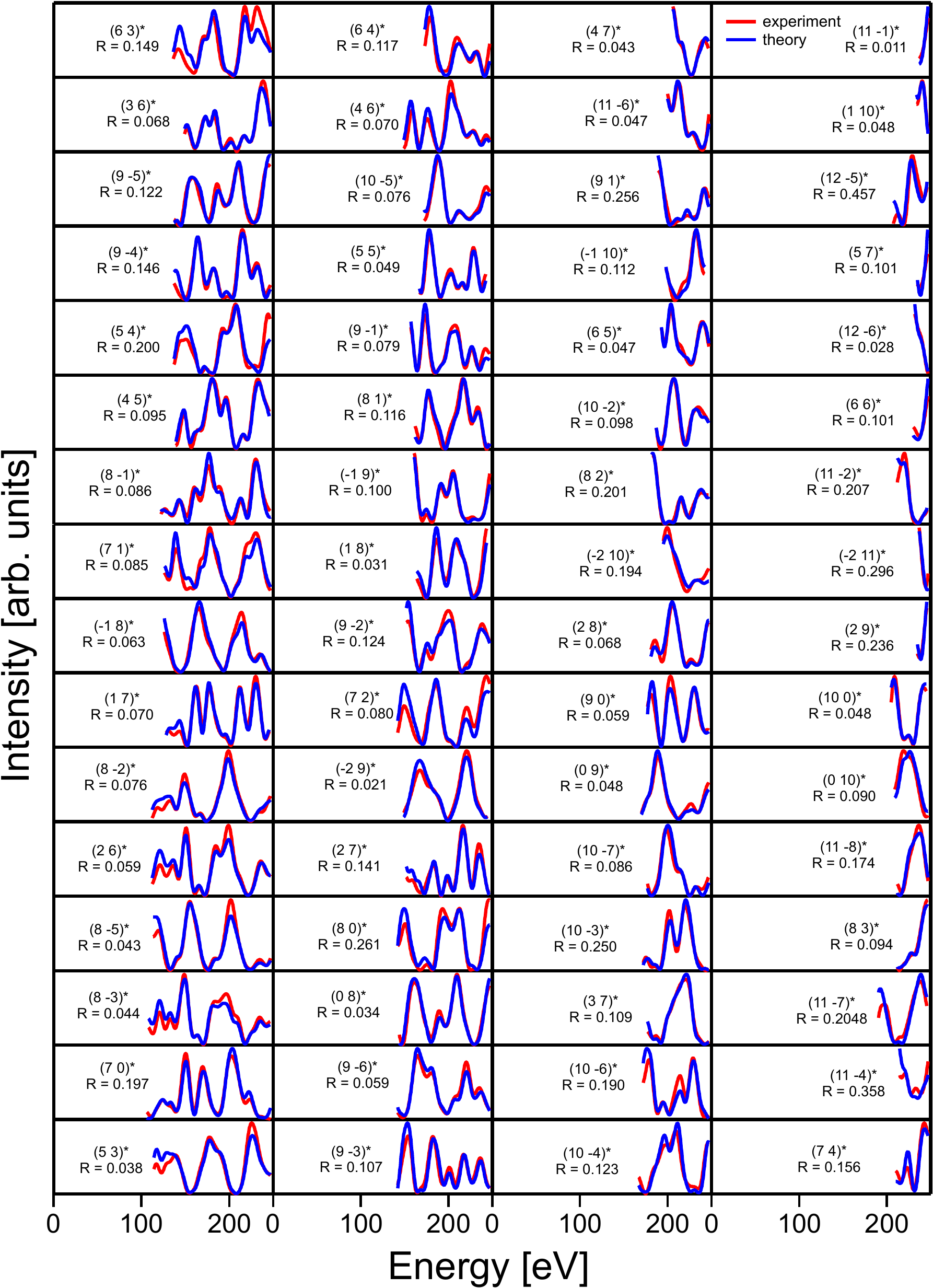}
\end{center}
\end{minipage}
\vspace{1cm}

\noindent Compilation of all 128 experimental and calculated bestfit
spectra entering the LEED-IV analysis. The beam labeling corresponds
to the reciprocal mesh of the ($\sqrt{13} \times
\sqrt{13}$)R13.9$^\circ$ superstructure.

\clearpage

\section*{Error curves for fitted atomic displacements}

\begin{minipage}[t]{1\textwidth}
\begin{center}
\includegraphics[width=0.84\textwidth]{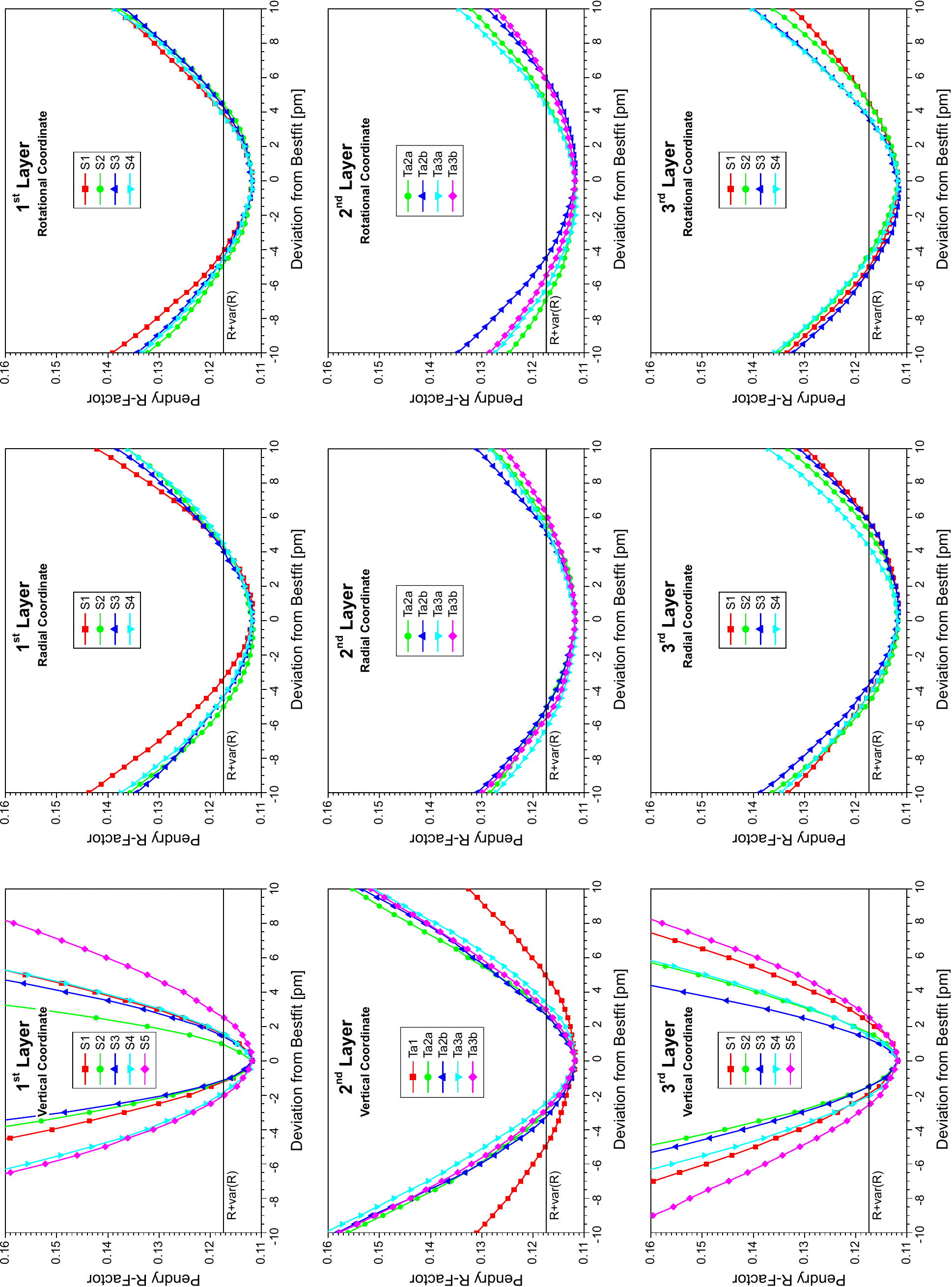}
\end{center}
\end{minipage}

\clearpage

\begin{minipage}[t]{1\textwidth}
\begin{center}
\includegraphics[width=0.84\textwidth]{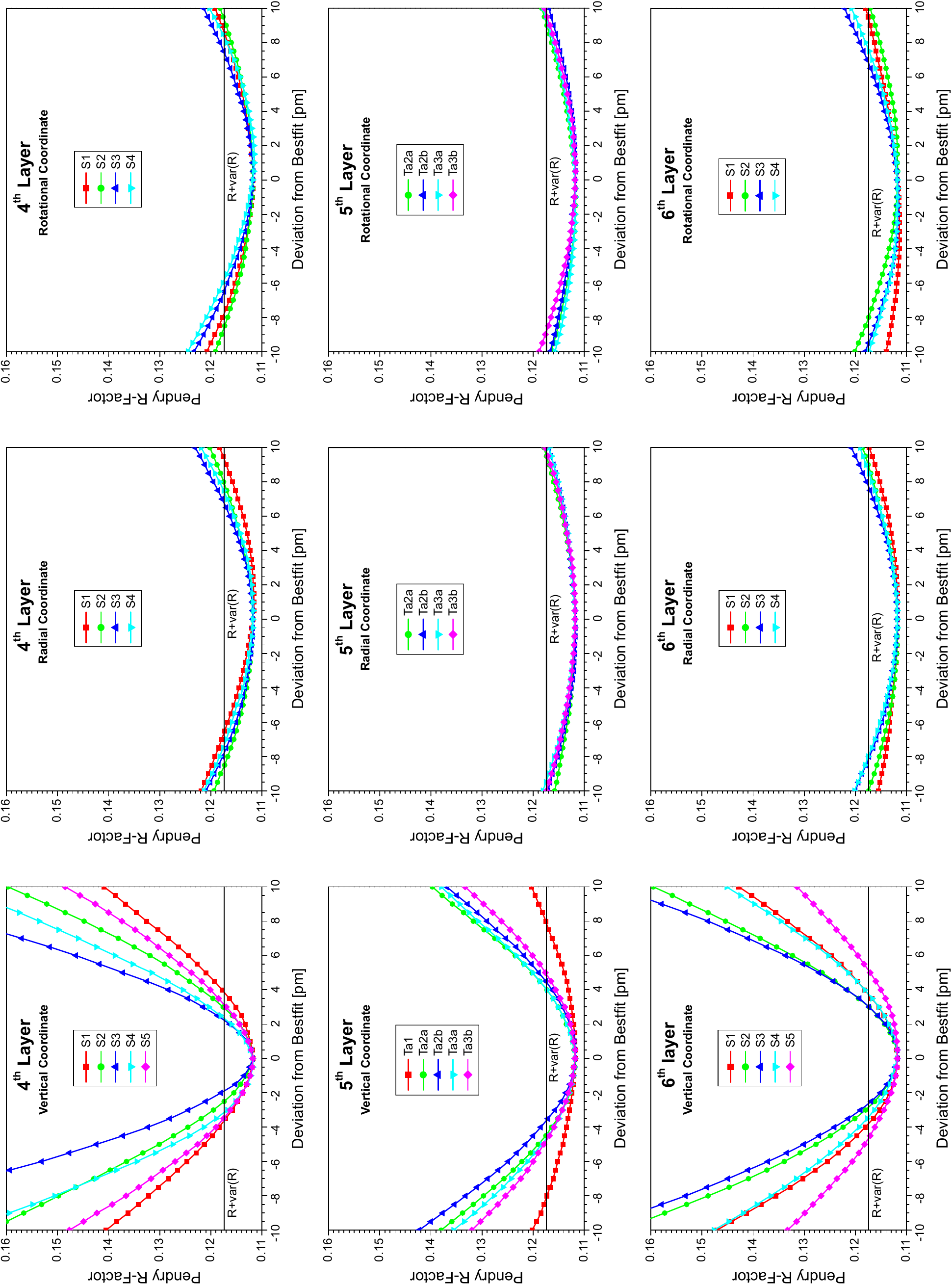}
\end{center}
\end{minipage}
\vspace{1cm}

\noindent R-factor variation as a function of displacement from the
bestfit value (``error curves'') for all site parameters of layers 1
-- 6 varied in the analysis. For reasons of computational time
saving the error analysis was not performed for azimuthally averaged
spectra but for one single azimuth of incidence only ($\Phi=0$).
Therefore, the error curves as well as the variance level are
vertically shifted by $\Delta R = 0.002$ compared to the bestfit.

\end{document}